\documentclass[sigconf]{acmart}
\usepackage{booktabs} 

\usepackage[english]{babel}
\usepackage{moresize}
\usepackage{amsmath}
\usepackage{algorithmic}
\usepackage{balance}
\usepackage{comment}
\usepackage{paralist}
\usepackage{bm}
\usepackage{pgfplots}
\usetikzlibrary{pgfplots.dateplot}

\usepackage{flushend}
\usepackage[english]{babel}
\usepackage[latin1]{inputenc}
\usepackage{mathrsfs}
\usepackage{graphicx}

\usepackage{amssymb}
\usepackage{amsfonts}
\usepackage{url}
\usepackage{longtable}
\usepackage{rotating}
\usepackage{multirow}
\usepackage{mathrsfs}
\usepackage{subfigure}
\usepackage{enumitem}
\usepackage[linesnumbered,algoruled,boxed,lined]{algorithm2e}
\usepackage{adjustbox}
\usepackage{hyperref}
\usepackage{pgfplots}
\usetikzlibrary{pgfplots.dateplot}
\usepackage{filecontents}
\definecolor{tblue}{RGB}{31,119,180}
\definecolor{torange}{RGB}{255,127,14}
\definecolor{tgreen}{RGB}{44,160,44}
\definecolor{tred}{RGB}{214,39,40}
\definecolor{tpurple}{RGB}{148,103,189}

\newcommand{\hide}[1]{} 

\newcommand{\ie}{\textit{i}.\textit{e}.}
\newcommand{\eg}{\textit{e}.\textit{g}.} 
\newcommand{\wrt}{\textit{w}.\textit{r}.\textit{t}}

\def\model{AutoCF}

\setcopyright{none}


\begin{document}
\fancyhead{}






\copyrightyear{2023}
\acmYear{2023}
\setcopyright{acmlicensed}\acmConference[WWW'23]{Proceedings of the ACM Web Conference 2023}{April 30-May 4, 2023}{Austin, TX, USA}
\acmBooktitle{Proceedings of the ACM Web Conference 2023 (WWW'23), April 30-May 4, 2023, Austin, TX, USA}
\acmPrice{15.00}
\acmDOI{10.1145/3543507.3583336}
\acmISBN{978-1-4503-9416-1/23/04}

\title{Automated Self-Supervised Learning for Recommendation}


\author{Lianghao Xia}
\affiliation{The University of Hong Kong}
\email{aka\_xia@foxmail.com}

\author{Chao Huang}
\authornote{Chao Huang is the corresponding author.}
\affiliation{The University of Hong Kong}
\email{chaohuang75@gmail.com}

\author{Chunzhen Huang}
\affiliation{Tencent}
\email{chunzhuang@tencent.com}

\author{Kangyi Lin}
\affiliation{Tencent}
\email{plancklin@tencent.com}

\author{Tao Yu}
\affiliation{The University of Hong Kong}
\email{tyu@cs.hku.hk}

\author{Ben Kao}
\affiliation{The University of Hong Kong}
\email{kao@cs.hku.hk}


\begin{abstract}
Graph neural networks (GNNs) have emerged as the state-of-the-art paradigm for collaborative filtering (CF). To improve the representation quality over limited labeled data, contrastive learning has attracted attention in recommendation and benefited graph-based CF model recently. However, the success of most contrastive methods heavily relies on manually generating effective contrastive views for heuristic-based data augmentation. This does not generalize across different datasets and downstream recommendation tasks, which is difficult to be adaptive for data augmentation and robust to noise perturbation. To fill this crucial gap, this work proposes a unified \underline{Auto}mated \underline{C}ollaborative \underline{F}iltering (\model) to automatically perform data augmentation for recommendation. Specifically, we focus on the generative self-supervised learning framework with a learnable augmentation paradigm that benefits the automated distillation of important self-supervised signals. To enhance the representation discrimination ability, our masked graph autoencoder is designed to aggregate global information during the augmentation via reconstructing the masked subgraph structures. Experiments and ablation studies are performed on several public datasets for recommending products, venues, and locations. Results demonstrate the superiority of \model\ against various baseline methods. We release the model implementation at \url{https://github.com/HKUDS/AutoCF}.
\end{abstract}

\maketitle

\section{Introduction}
\label{sec:intro}

Recommender systems, which aim to suggest items to users by learning their personalized interests, have provided essential web services (\eg, E-commerce sites, online reviews, and advertising platforms) to alleviate the information overload problem~\cite{wu2022survey}. The core part of recommenders lies in effective modeling of user preference on various items based on observed historical interactions.

To date, various types of collaborative filtering (CF) techniques have been proposed to project users and items into latent embedding space, such as matrix factorization~\cite{koren2009matrix}, autoencoder~\cite{sedhain2015autorec}, attention mechanism~\cite{chen2017attentive}. In the context of graph learning, graph neural networks (GNNs) have emerged as the state-of-the-art frameworks for collaborative filtering (CF) with the modeling of high-order connectivity among users and items, such as GCMC~\cite{berg2017graph}, NGCF~\cite{wang2019neural}, LightGCN~\cite{he2020lightgcn} and GCCF~\cite{chen2020revisiting}. These recommender systems are proposed to perform the recursive message passing over the generated user-item interaction graph. However, the success of such methods largely relies on sufficient high quality labels, and cannot produce accurate user and item representations when the observed labeled data is scarce and noisy~\cite{wu2021self}. Recently, contrastive self-supervised learning (SSL) has achieved promising results in generating representations with small labeled data with yielding auxiliary self-supervision signals in Computer Vision~\cite{he2020momentum} and Nature Language Processing~\cite{pan2021contrastive}. Motivated by this, recent recommendation studies propose to address the limitation of heavy label reliance in current GNN-based recommendation models based on various contrastive learning techniques for data augmentation.


According to the ways in which contrastive views are generated for collaborative filtering, recent contrastive learning models can be broadly categorized into different aspects: i) \emph{Structure-Level Augmentation}: Models such as SGL~\cite{wu2021self} and DCL~\cite{liu2021contrastive} employ random node and edge dropout to create contrastive views based on graph topology augmentation. ii) \emph{Feature-Level Augmentation}: SLRec~\cite{yao2021self} generates contrasts between feature vectors that have been corrupted by random noise for augmentation purposes. iii) \emph{Local-Global Augmentation}: This line aims to reconcile local user and item embeddings with global information by performing local-global contrastive learning. To achieve this goal, various information aggregators are used to produce global-level embeddings, including hypergraph-enhanced message fusion in HCCF~\cite{xia2022hypergraph} and EM algorithm-based node clustering in NCL~\cite{lin2022improving}.

However, current contrastive learning recommenders heavily rely on manually generated contrastive views for heuristic-based data augmentation. The effectiveness of these approaches is guaranteed only when the contrastive views are properly specified for different datasets and tasks. In diverse recommendation scenarios, accurate generation of contrastive representation views is very challenging and manually performing data augmentation may unavoidably involve noisy and irrelevant information for SSL.

\begin{figure}[t]
    \centering
    \subfigure[Impact of data noise]{
        \includegraphics[width=0.3\columnwidth]{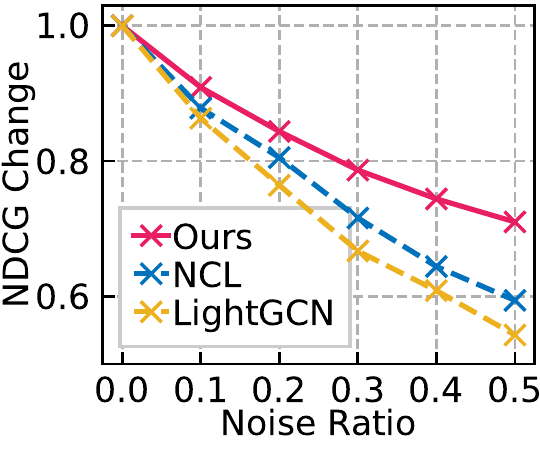}
    }
    \subfigure[Impact of long-tail dist]{
        \includegraphics[width=0.3\columnwidth]{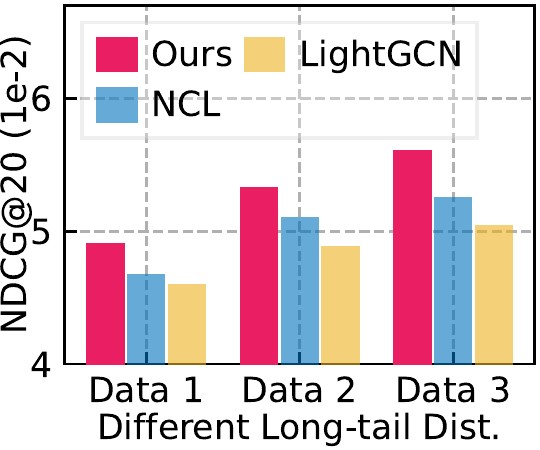}
    }
    \subfigure[Dist. of datasets]{
        \includegraphics[width=0.3\columnwidth]{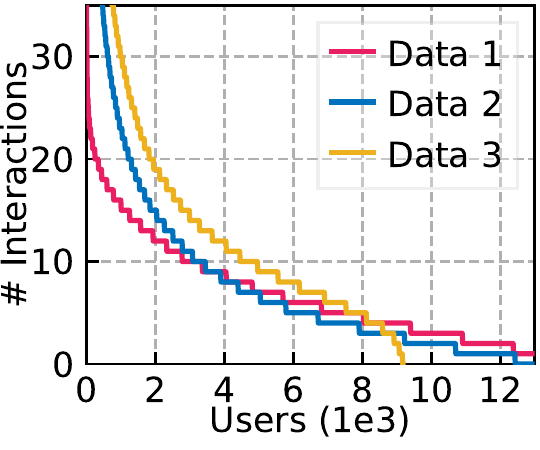}
    }
    \vspace{-0.12in}
    \caption{The influence of data noise and long-tail distributions on the prediction accuracy of different methods.}
    \vspace{-0.1in}
    \label{fig:intro}
    \Description{A figure showing the performance of \model\ and representative baselines with respect to noise ratio and different data distributions. \model\ consistently outperforms the baselines. Specially, \model\ is shown to be significantly more robust to high-ratio of noises.}
\end{figure}

From the perspective of graph structure-level and local-global augmentation, the self-discrinimation through random node/edge dropout operations may lead to i) losing important structural information (\eg, limited interactions of inactive users); and ii) keeping the noisy data (\eg, misclick behaviors or popularity bias) for contrasting samples. To have a better understanding of the aforementioned limitations in current methods, Figure~\ref{fig:intro} presents the performance comparison of different methods against data noise and long-tail distributions. Specifically, we contaminate the training set with different ratios of adversarial user-item interactions as noisy examples. In addition, we randomly sample three datasets from Gowalla with different long-tail distributions (shown in Figure~\ref{fig:intro} (c)). The superior performance of our \model\ indicates that existing state-of-the-art SSL recommendation models can hardly learn high-quality user (item) representations when faced with data noise and long-tail issues. The performance of compared current contrastive methods (\eg, NCL~\cite{lin2022improving}) vary greatly with different recommendation data distributions.

Therefore, current contrastive recommender systems are still vulnerable to the quality of the augmented supervision signals, due to their handcrafted strategies of non-adaptive contrastive view generation. With the consideration of limitations in existing work, we believe it is essential to design a unified SSL-based recommendation model, which not only can distill self-supervised signals for effective data augmentation, but also relieve human efforts to manually in generating self-supervision signals or defining augmentation strategies. Towards this end, an interesting question may be raised: \emph{is there a principled way to automatically distill the important self-supervision signals for adaptive augmentation?}

Inspired by the emerging success of generative self-supervised learning in vision learner~\cite{he2022masked} with the reconstruction objectives, we propose an automated framework for self-supervised augmentation in graph-based CF paradigm via a masked graph auto-encoder architecture, to explore the following questions for model design.
\begin{itemize}[leftmargin=*]
\item \textbf{Q1}: How to automatically distill self-supervised signals which are more beneficial for the recommendation objective?
\item \textbf{Q2}: How to enable the learnable graph augmentation by well preserving informative collaborative relationships?
\item \textbf{Q3}: How to design the graph auto-encoder framework in global information aggregation for the adaptive graph reconstruction.
\end{itemize}

\noindent \textbf{Contributions}. To tackle these challenges, we propose a new \underline{Auto}mated \underline{C}ollaborative \underline{F}litering (\model) which is capable of distilling the graph structure-adaptive self-supervised signals for advancing the graph-based CF framework. In particular, we design a learnable masking function to automatically identify the important centric nodes for reconstruction-based data augmentation. In the mask learning stage, the node-specific subgraph semantic relatedness will be considered to accurately preserve the graph-based collaborative relationships. Additionally, we propose a new masked graph autoencoder in which the key ingredient is a graph neural encoder that captures the global collaborative relationships for reconstructing the masked user-item subgraph structures.


In summary, our work makes the following contributions:
\begin{itemize}[leftmargin=*]

\item Investigate the drawbacks of existing contrastive GNN-based recommendation methods with the non-adaptive self-supervised augmentation and weak robustness against noise perturbation. \\\vspace{-0.12in}

\item Propose an automated self-supervised learning model \model, in which a learning to mask paradigm is designed to perform data augmentation with structure-adaptive self-supervision. In addition, the automated mask generator is integrated with graph masked autoencoder to enhance user representations with SSL. \\\vspace{-0.12in}

\item Demonstrate the significant improvements that \model\ achieves over state-of-the-art GNN-based and SSL-enhanced recommenders, including some recent strong baselines NCL and HCCF.

\end{itemize}


\section{Graph Contrastive Learning for Collaborative Filtering}
\label{sec:model}

\noindent \textbf{Recap of Graph-based Collaborative Filtering}. In general, the input data for recommender systems involve an user set $\mathcal{U}=\{u\}$, an item set $\mathcal{I}=\{i\}$ and observed user-item relations represented by an interaction matrix $\textbf{A}\in\mathbb{R}^{|\mathcal{U}|\times |\mathcal{I}|}$. In $\textbf{A}$, each element $a_{u,i}=1$ if user $u$ has adopted item $i$ before, $a_{u,i}=0$ given the non-interaction between user $u$ and item $i$. To improve the user-item interaction modeling with high-order connectivity, graph-based collaborative filtering models have shown their effectiveness through the recursive message propagation over the constructed interaction graph $\mathcal{G}=\{\mathcal{U}, \mathcal{I}, \mathcal{E}\}$ between $u$ and $i$ for embedding refinement :
\begin{align}
\textbf{h}_u^l \leftarrow \textbf{Aggre}_{i\in \mathcal{N}_u} \Big ( \textbf{Const} ( \textbf{h}_i^{l-1}, \textbf{h}_u^{l-1}, e_{u,i}) \Big ) \nonumber\\
\textbf{h}_i^l \leftarrow \textbf{Aggre}_{u\in \mathcal{N}_i} \Big ( \textbf{Const} ( \textbf{h}_u^{l-1}, \textbf{h}_i^{l-1}, e_{i,u}) \Big )
\end{align}
\noindent where in $\mathcal{G}$, users and items are connected through their interaction edges ($e_{u,i}$) based on the interaction matrix $\textbf{A}$. $\textbf{h}_u^{l-1}$ and $\textbf{h}_i^{l-1}$ denote the embedding of user $u$ and item $i$ at the $(l-1)$-th graph neural layer. The core of graph-based CF paradigm contains: i) the message construction function $\textbf{Const}(\cdot)$ which extracts the feature information from the connected users or items; ii) the information aggregation function $\textbf{Aggre}(\cdot)$ for gathering the embeddings from the neigboring nodes $i\in \mathcal{N}_u$ or $u\in \mathcal{N}_i$, using different operators, such as mean-pooling, summation, or attentive combination.\\\vspace{-0.12in}

\noindent \textbf{Contrastive Augmentation on Interaction Graph}. Inspired by the recent success of contrastive learning~\cite{tian2020makes,khosla2020supervised}, several recent studies propose to perform graph contrastive learning (GCL) over the user-item interaction data with various data augmentation strategies. In general, contrastive learning-enhanced recommender systems aim to reach the alignment between the generated contrasting representation views, so as to inject the auxiliary self-supervised objective $\mathcal{L}_\text{ssl}$ into the recommendation loss $\mathcal{L}_\text{rec}$ with the joint model optimization for estimating the $u-i$ interaction likelihood:
\begin{align}
    \hat{y}_{u,i}=f(\mathcal{G}; \mathbf{\Theta})[u,i];~~~~~\min\limits_{\mathbf{\Theta}} \mathcal{L}_\text{rec}(f, \mathcal{G}) + \mathcal{L}_\text{ssl}(f, \mathcal{G};\varphi)
\end{align}

However, the success of most state-of-the-art GCL-based recommendation solutions largely relies on the careful design of contrastive views and handcrafted augmentation strategies $\varphi$. To address this limitation, this work aims to design graph augmentation scheme that enables the automated self-supervision signal generation for adaptive data augmentation in recommendation.




\section{Methodology}
\label{sec:solution}

\begin{figure*}
    \centering
    \includegraphics[width=0.98\textwidth]{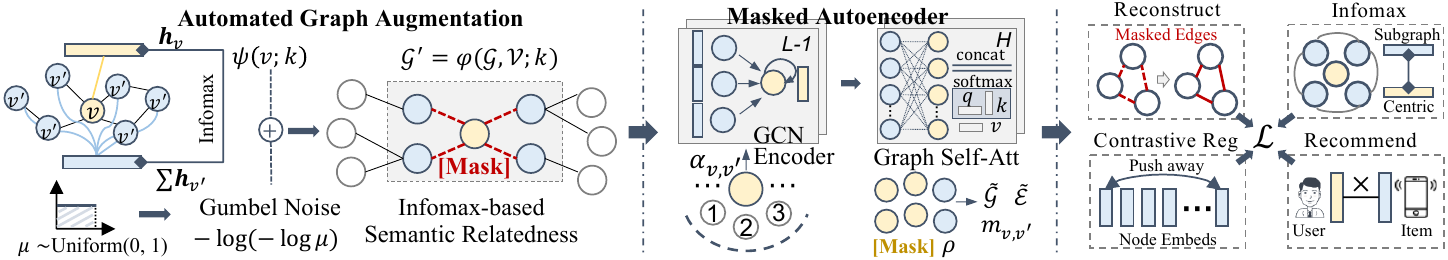}
    \vspace{-0.1in}
    \caption{The framework of \model\ is composed of: i) the adaptive graph structure augmentation based on infomax relatedness; ii) the masked graph autoencoder based on the graph self-attention architecture; iii) the self-augmented training paradigm including graph structure reconstruction, node-subgraph mutual information maximization, and contrastive learning.}
    \Description{A figure illustrating the framework of the proposed \model\ model, which is composed of three parts: the adaptive graph augmentation, the masked autoencoder, and the self-augmented training paradigm.}
    \label{fig:framework}
    \vspace{-0.1in}
\end{figure*}


This section elaborates the technical details of adaptive masked graph autoencoder--\model--to enhance user representation with the automated distillation of self-supervision signals based on graph neural networks. We present the architecture of \model\ in Figure~\ref{fig:framework}.

\vspace{-0.05in}
\subsection{Automated Graph Augmentation}


\subsubsection{\bf Learning to Mask Paradigm}
To automatically distill the reconstructed self-supervised signals over the graph-structured interaction data $\mathcal{G}$, we propose to reconsutrct the masked user-item interaction edges adaptively for benefiting the modeling of high-order graph collobrative relationships. The core idea of our learnable mask scheme is to first identify the centric nodes in $\mathcal{G}$ and then mask the informative interaction edges based on their subgraph structual information. In general, the learnable graph-based interaction mask function $\varphi(\cdot)$ in \model\ can be formalized:
\begin{align}
    \label{eq:masking}
    \varphi(\mathcal{G}, \mathcal{V}; k) = \left\{\mathcal{U}, \mathcal{I}, \mathcal{V}, \mathcal{E} \backslash \ \{(v_1, v_2)|v_1, v_2\in\mathcal{N}^k_v, v\in\mathcal{V}\}\right\}
\end{align}


\noindent where $\mathcal{V}$ is the set of centric nodes in $\varphi(\cdot)$. $\mathcal{N}_v^k$ denotes the set of neighborhood nodes of $v$ within $k$ hops. Here, $k$ is a hyperparameter to control the injection of high-order connectivity during the reconstruction-based self-augmentation. $(v_1,v_2)$ denotes the existing edge masked through the set subtraction operation ``$\backslash$'', in the sampled subgraph structure given the centric node $v\in\mathcal{V}$.



\subsubsection{\bf Infomax-based Semantic Relatedness} Towards this end, motivated by the mutual information encoding over graphs~\cite{velickovic2019deep,sun2019infograph}, we leverage the mutual information to measure the semantic relatedness between the node-level embedding and subgraph-level representation, based on the high-order graph collaborative relations among users and items. Formally, the subgraph semantic relatedness score $s_v$ of node $v$ can be derived as follows:
\begin{align}
\label{eq:infomax}
s_v =\psi(v;k)
= \text{sigm}(\textbf{h}_v^\top \sum_{v'\in\mathcal{N}_v^k} \textbf{h}_{v'} / (|\mathcal{N}_v^k|\cdot \|\textbf{h}_v\|\cdot \|\textbf{h}_{v'}\|))
\end{align}
\noindent where $\textbf{h}_v, \textbf{h}_{v'}\in\mathbb{R}^d$ denote randomly-initialized user/item embeddings. $\text{sigm}(\cdot)$ dentoes the sigmoid activation.
The subgraph-level representation is generated by aggregating the embeddings of all nodes ($v'\in\mathcal{N}_v^k$) contained in this $k$-order subgraph except the centric node $v$, using readout functions, such as mean-pooling or summation~\cite{velickovic2019deep}. The larger semantic relatedness score indicates not only the higher structural consistency between the target user and his/her graph dependent nodes (users and items), but also the lower percentage of topological information noise in the sampled subgraph. For example, an outlier user with many item misclicks will lead to lower structural consistency with others with respect to their collaborative relationships of interaction data.

\subsubsection{\bf Learning to Mask with Gumbel Distribution}
\label{sec:infomax_learning}
Given the designed learnable masking function $\varphi(\mathcal{G}, \mathcal{V}; k)$, our \model\ is capable of automatically generating the self-supervised reconstruction signals of user-item interactions, which offers adaptive data augmentation. To improve the robutness of our learning to mask paradigm $\varphi(\mathcal{G}, \mathcal{V}; k)$, we inject the Gumbel-distributed noises~\cite{jang2017categorical} into the deriviation of node-specific mask probability $\psi(v;k)$:
\begin{align}
    \label{eq:gumbel}
    {\psi}'(v;k) = \log{\psi}(v;k)-\log(-\log(\mu));~~~~~ \mu\sim\text{Uniform}(0, 1)
\end{align}
\noindent Based on the estimated mask probabilities ${\psi}'(v;k)$ of all nodes in $v\in \mathcal{V}$, we generate a set of $S$ centric nodes by selecting the top-ranked user and item nodes in terms of the learned mask probabilities. To supercharge our \model\ with learnable data augmentation, we further inject the SSL signals via the subgraph mutual information maximization using the following infomax-based opimized objective $\mathcal{L}_\text{InfoM}=-\sum \nolimits_{v\in\mathcal{U}\cup\mathcal{I}} \psi(v; k)$.



\vspace{-0.05in}
\subsection{Masked Graph Autoencoder}
The goal of \model\ is to augment the graph-based collaborative filtering with a reconstruction learning task over the masked user-item interaction edges in graph $\mathcal{G}$. After applying our learning to mask paradigm, we feed the augmented graph with the masked edges into our developed graph autoencoder framework. In particular, \model\ proposes to leverage graph convolutional network, which is widely used in previous recommender systems~\cite{he2020lightgcn}, as the encoder to incorporate graph structural information into user and item node embeddings. To alleviate the over-smoothing issue in GNNs and supercharge \model\ with the global information aggregation, we adopt graph self-attention as the decoder to bridge the encoder and the auxiliary self-supervised reconstruction task.


\vspace{-0.05in}
\subsubsection{\bf Graph Convolution-based Structure Encoder}
Recent studies (\eg, SGL~\cite{wu2021self} and LightGCN~\cite{he2020lightgcn}) have demontrated the effectiveness of lightweight graph convolution-based message passing for encoding the structual information. Motivated by this, given the corrupted interaction graph $\mathcal{G}'=\varphi(\mathcal{G}, \mathcal{V};k)$ output from our mask function $\varphi(\mathcal{G}, \mathcal{V}; k)$, our encoder in \model\ is built over the following embedding propagation schema:
\begin{align}
    \label{eq:gcn}
    \textbf{h}^{l+1}_v= \alpha_{v,v}\cdot \textbf{h}^l_v + \sum_{v'\in\mathcal{N}'_v} \alpha_{v,v'} \cdot \textbf{h}^{l}_{v'}; ~~~~~
    \alpha_{v,v'} = 1 / \sqrt{|\mathcal{N}'_v||\mathcal{N}'_{v'}|}
\end{align}
\noindent where $\textbf{h}_v^{l}, \textbf{h}_v^{l+1} \in\mathbb{R}^d$ denotes the node embeddings of node $v$ in the $l$-th and the $(l+1)$-th graph neural layers, respectively. The scalar $\alpha_{v,v'}$ serves as the normalization weight for node pair $(v,v')$, which is calculated based on the node degree $|\mathcal{N}'_v|, |\mathcal{N}'_{v'}|$. $\mathcal{N}'_v$ denotes the neighborhood set of node $v$ in the augmented graph $\mathcal{G}'$. During our embedding passing process, \model\ applies the residual connections to enable the self-propagation in the last graph layer, in order to alleviate the gradient vanishing issue~\cite{he2016deep}.

\vspace{-0.05in}
\subsubsection{\bf Graph Self-Attention Decoder}
\label{sec:graphTrans}
Although the graph convolutional encoder allows us to capture the user-item interaction graph structure, the over-smoothing problem will become a natural effect with increasing the graph propagation layers~\cite{chen2020simple,chen2020measuring}. To alleviate this limitation, we design the graph self-attention module as the decoder in \model\ for the self-supervised structure reconstruction, with the global self-attention for long-range information aggregation, rather than the localized convolutional fusion.



While the attention-based information aggregation addresses the limitation of localized GCN with limited receptive field, the high computational complexity limits its feasibility to perform message passing over the large number of graph user and item nodes. Inspired by the efficient transformer design in encoding long sequences~\cite{kitaev2020reformer} and high-resolution images~\cite{zamir2022restormer}, we propose to conduct the pairwise relation learning over a subset of nodes with the emphasis on the masked subgraph structure. By doing so, we can not only improve the efficiency of our graph self-attention decoder, but also further capture the high-order structural information of centric nodes $\mathcal{V}$ with high subgraph semantic relatedness.
Specifically, we firstly define a vertex set $\bar{\mathcal{V}}$ to include the vertices from all the masked subgraphs. Given $\bar{\mathcal{V}}$, a subset $\tilde{\mathcal{V}}$ of nodes will be added from the remaining nodes ($(\mathcal{U} \cup \mathcal{I}) \backslash \bar{\mathcal{V}}$). Then, node pairs $\bar{\mathcal{E}}$ are then selected from the union node set, and combined with the edges in $\mathcal{G}'$ using the following formulas:
\begin{align}
    \label{eq:gatSample}
    \tilde{\mathcal{G}} &= \{\mathcal{U}, \mathcal{V}, \tilde{\mathcal{E}}=\bar{\mathcal{E}}\cup\mathcal{E}'\};~~~~~ \bar{\mathcal{E}}=\{(v_1, v_2)|v_1, v_2 \in \bar{\mathcal{V}}\cup\tilde{\mathcal{V}}\} \nonumber\\
    &\text{s.t.} ~~|\bar{\mathcal{E}}| = |\mathcal{E}'|, ~~ |\bar{\mathcal{V}}\cup\tilde{\mathcal{V}}|=\rho\cdot(|\mathcal{U}|+|\mathcal{I}|)
\end{align}
\noindent where $\mathcal{E}'$ denotes the edge set of the augmented graph $\mathcal{G}'=\varphi(\mathcal{G},\mathcal{V};k)$. $\rho$ is a hyperparameter to control the ratio of the node set. Given the constructed node pairs for global self-attention aggregation, the graph attention-based message passing is presented:
\begin{align}
    \label{eq:graphTrans}
    &\textbf{h}_v^{l+1} = \sum_{v'} \mathop{\Bigm|\Bigm|}\limits_{h=1}^H m_{v,v'} \beta_{v, v'}^h  \textbf{W}_\text{V}^h \textbf{h}_{v'}^l;~~~ m_{v,v'}=\left\{
    \begin{aligned}
    &1~~\text{if}~(v,v')\in\tilde{\mathcal{E}}\\
    &0~~\text{otherwise}
    \end{aligned}\right.
    \nonumber\\
    & \beta^h_{v,v'} = \frac{\exp \bar{\beta}^h_{v,v'}}{\sum_{v'}\exp\bar{\beta}^h_{v,v'} };~~~~~~~~
    \bar{\beta}^h_{v,v'} = \frac{(\textbf{W}_\text{Q}^h \cdot \textbf{h}_v^{l})^\top \cdot (\textbf{W}_\text{K}^h \cdot \textbf{h}_{v'}^{l})}{\sqrt{d/H}}
\end{align}
\noindent where $H$ denotes the number of attention heads (indexed by $h$). $m_{v,v'}$ is the binary indicator to decide whether to calculate the attentive relations between node $v$ and $v'$. $\beta_{v,v'}^h$ denotes the attention weight for node pair $(v,v')$ \wrt\ the $h$-th head representation space. $\textbf{W}_\text{Q}^h, \textbf{W}_\text{K}^h, \textbf{W}_\text{V}^h \in \mathbb{R}^{d/H\times d}$ denotes the query, the key, and the value embedding projection for the $h$-th head, respectively.\\\vspace{-0.12in}

Different from conventional graph auto-encoders~\cite{hasanzadeh2019semi,tang2019correlated}, our \model\ aims to recover the masked interaction graph structure by learning to discover the masked user-item edges. With the encoded layer-specific user/item representations $\textbf{h}_u^l$ and $\textbf{h}_i^l$, the overall embeddings are generated through layer-wise aggregation. Formally, the reconstruction phase over the masked graph structures is:
\begin{align}
    \label{eq:pred}
    \mathcal{L}_\text{recon} = -\sum_{(v,v')\in\mathcal{E}\backslash\mathcal{E}'} \hat{\textbf{h}}_v^\top \cdot \hat{\textbf{h}}_{v'};~~~~~
    \hat{\textbf{h}}_v = \sum_{l=0}^L \textbf{h}_v^l
\end{align}


\vspace{-0.05in}
\subsection{Model Training}
In our model training stage, we further introduce a contrastive training strategy to enhance the representation discrinimation ability with uniformly-distributed user/item embeddings, so as to better preserve the unique preference of users in latent space. Inspired by the InfoNCE-based augmentation in~\cite{wu2021self}, in the learning process of \model, we propose to generate more uniform user and item embeddings with the regularization for user-item, user-user, item-item pairs, for improving the embedding discrinimation ability and further alleviating the over-smoothing effect. The loss function $\mathcal{L}_\text{ssl}$ with the augmented self-supervised learning objectives is:
\begin{align}
    \label{eq:ssl}
    \mathcal{L}_\text{ssl} = &\sum_{u\in\mathcal{U}}  \log \sum_{i\in\mathcal{I}} \exp \hat{\textbf{h}}_u^\top \hat{\textbf{h}}_i + \sum_{u\in\mathcal{U}} \log\sum_{u'\in\mathcal{U}} \exp \hat{\textbf{h}}_u^\top \hat{\textbf{h}}_{u'}\nonumber\\
    &+ \sum_{i\in\mathcal{I}}\log\sum_{i'\in\mathcal{I}} \exp \hat{\textbf{h}}_i^\top \hat{\textbf{h}}_{i'} +\mathcal{L}_\text{InfoM} + \mathcal{L}_\text{recon}
\end{align}

To perform the learning process with the main recommendation task and augmented SSL optimized objectives, we define our joint loss function for model optimization as follows:
\begin{align}
    \label{eq:loss}
    \mathcal{L}=-\sum_{(u,i)\in\mathcal{E}}\hat{\textbf{h}}_u^\top \cdot \hat{\textbf{h}}_i + \lambda_1 \cdot \mathcal{L}_\text{ssl} + \lambda_2\cdot\|\mathbf{\Theta}\|_\text{F}^2
\end{align}
\noindent $\lambda_1$ and $\lambda_2$ control the regularization strength of the distilled self-supervised signals and weight-decay constrain term.

\vspace{-0.05in}
\subsection{Theoretical and Complexity Analysis}
\subsubsection{\bf Noise Filtering via Infomax}
In this section, we give a theoretical discussion on the benefits of our infomax-based semantic relatedness in alleviating noisy signals for SSL-based augmentation. By avoiding involving noise in our adaptive graph augmentation using infomax, the masked autoencoding task better benefits the downstream recommendation task with noise-less gradients. Detailed derivations are presented in Section~\ref{sec:analysis}. In brief, we first analyze the source of noises by giving the gradient of $\mathcal{L}_\text{recon}$ over the embedding $\textbf{h}_{v_1}$ of a node $v_1$ in the masked subgraph as:
\begin{align}
    \label{eq:grad}
     \frac{\partial \mathcal{L}_\text{recon}}{\partial\textbf{h}_{v_1}} = -\sum_{(v_1,v_2)\in\mathcal{E}\backslash\mathcal{E}'} \textbf{h}_{v_2} + \sum_{m_{v_2,v'}=1}\beta'_{v_2,v'} \textbf{W}'_\text{V} \textbf{h}_{v'}
\end{align}
\noindent where node $v_2$ is adjacent to $v_1$, $v'$ is an arbitrary node, most-probably from the masked subgraph due to our mask-dependent sampling strategy (Section~\ref{sec:graphTrans}). By inspecting the gradient in Eq~\ref{eq:grad}, we find that the second term introduces noise if $v'$ is semantically less-relevant to $v_1$. Specifically, although the weight $\beta'_{v_2, v'}$ can reduce the influence of irrelevant $v'$, it is normalized using softmax with high temperature, which in practical learning process, is prone to assigning non-neligable weights to less-relevant $v'$ when a large portion of nodes in the masked subgraph are less-relevant.

To avoid the above situation, \model\ learns to mask noise-less subgraphs by referring to the infomax-based relatedness $s_v$. For a centric node $v$, $s_v$ not only measures the relatedness between $v$ and its neighboring nodes, but also restricts the lowerbound of semantic relatedness between any pair of nodes in the subgraph, as follows:
\begin{align}
    \cos(v_1, v') > \cos(v, v_1) + \cos(v, v') - 1
\end{align}
As subgraphs with larger $s_v$ have larger center-neighbor similarity $\cos(v, v_1)$ and $\cos(v, v')$, larger $s_v$ also indicates higher lowerbound for the semantic relatedeness between any nodes $(v_1, v')$ in the subgraph. In other words, by masking subgraphs with larger infomax-based relatedness, \model\ reduces the likelihood of introducing noisy gradient in the self-augmented reconstruction task.

\vspace{-0.05in}
\subsubsection{\bf Model Complexity Analysis}
We analyze the time complexity of our \model\ from different components. i) In the learning to mask paradigm, the subgraph-level embedding generation process takes $\mathcal{O}(k\times|\mathcal{E}|\times d)$ complexity for maintainaning a multi-order node intersection set. Here, $|\mathcal{E}|$ denotes the size of the edge set and $d$ is the latent embedding dimensionality. Additionally, the masked edge detection takes $\mathcal{O}(|\mathcal{E}|\times |\mathcal{V}_{k'}|)$ time, where $\mathcal{V}_{k'}$ denotes the set of centric nodes for the $k'$-th iteration ($\mathcal{V}_0=\mathcal{V}$, $1\leq k' \leq k$). ii) In the masked graph autoencoder component, the most time-consuming part is the graph self-attention which takes $\mathcal{O}(|\mathcal{E}|\times d^2)$ complexity. iii) In the model training phase, contrastive augmentation takes $\mathcal{O}(B\times (|\mathcal{U}|+|\mathcal{I}|)\times d)$ with $B$ denoting the batch size. In summary, our \model\ can achieve comparable overall complexity compared with state-of-the-art graph contrastive recommendation models, such as HCCF~\cite{xia2022hypergraph} and SGL~\cite{wu2021self}.

\section{Evaluation}
\label{sec:eval}


In this section, experiments are performed to validate \model's advantage and answer the following key research questions:
\begin{itemize}[leftmargin=*]
\item \textbf{RQ1}: How effective is our proposed \model\ method perform against various recommendation techniques?
\item \textbf{RQ2}: How effective are the main components of our \model?
\item \textbf{RQ3}: How is the sensitivity of the hyperparameters in \model?
\item \textbf{RQ4}: How efficient is our \model\ compared with baselines?
\item \textbf{RQ5}: How is the interpretation ability of the proposed \model?
\end{itemize}

\begin{table}[]
    \centering
    \caption{Statistics of the experimental datasets.}
    \label{tab:datasets}
    \small
    \vspace{-0.12in}
    \begin{tabular}{|c|cccc|}
        \hline
        Dataset & \# Users & \# Items & \# Interactions & Interaction Density \\
        \hline
        \hline
        Gowalla & 25,557 & 19,747 & 294,983 & $5.85\times 10^{-4}$\\
        Yelp & 42,712 & 26,822 & 182,357 & $1.59\times 10^{-4}$\\
        Amazon & 76,469 & 83,761 & 966,680 & $1.51\times 10^{-4}$\\
        \hline
    \end{tabular}
    \vspace{-0.1in}
    \Description{A table showing the statistics of the Gowalla data (25557 users, 19747 items, 294983 interactions), the Yelp data (42712 users, 26822 items, 182357 interactions), and the Amazon data (76469 users, 83761 items, 966680 interactions).}
\end{table}

\vspace{-0.1in}
\subsection{Experimental Settings}
\subsubsection{\bf Datasets and Evaluation Protocols}
We use three widely-used datasets, including Gowalla, Yelp and Amazon to evaluate the performance of different recommendation tasks for locations, venues and online products. \textbf{Gowalla}. This dataset is collected from a location-based service to record the check-in behaviors between users and different locations from Jan, 2010 to Jun, 2010. \textbf{Yelp}. This dataset contains user-venue rating interaction from Jan, 2018 to Jun, 2018 on Yelp platform. \textbf{Amazon}. This is another benchmark dataset in evaluating recommender systems. We use the version of user implicit feedback over the book category of items.


In our experiments, 70\% of the observed interactions are randomly sampled to generate the training set. In the remaining 30\% data, 5\% and 25\% percentage of interactions are used for validation and testing. Following the same settings in~\cite{wu2021self,xia2022hypergraph}, we adopt the all-rank evaluation protocol~\cite{krichene2022sampled} to measure the item recommendation accuracy for each user. Two representative evaluation metrics \emph{Recall@N} and \emph{NDCG@N} are used to evaluate all methods.


\begin{table*}
\vspace{-0.1in}
\caption{Performance comparison on Gowalla, Yelp, and Amazon datasets in terms of \textit{Recall} and \textit{NDCG}.}
\vspace{-0.15in}
\centering
\footnotesize
\setlength{\tabcolsep}{1.2mm}
\begin{tabular}{|c|c|c|c|c|c|c|c|c|c|c|c|c|c|c|c|c|c|l|}
\hline
Data & Metric & BiasMF & NCF & AutoR & PinSage & STGCN & GCMC & NGCF & GCCF & LightGCN & DGCF & SLRec & NCL & SGL & HCCF & \emph{\model} & p-val.\\
\hline
\multirow{4}{*}{Gowalla}
&Recall@20 & 0.0867 & 0.1019 & 0.1477 & 0.1235 & 0.1574 & 0.1863 & 0.1757 & 0.2012 & 0.2230 & 0.2055 & 0.2001 & 0.2283 & 0.2332 & 0.2293 & \textbf{0.2538} & $1.3e^{-10}$\\
&NDCG@20 & 0.0579 & 0.0674 & 0.0690 & 0.0809 & 0.1042 & 0.1151 & 0.1135 & 0.1282 & 0.1433 & 0.1312 & 0.1298 & 0.1478 & 0.1509 & 0.1482 & \textbf{0.1645} & $4.9e^{-12}$\\
\cline{2-18}
&Recall@40 & 0.1269 & 0.1563 & 0.2511 & 0.1882 & 0.2318 & 0.2627 & 0.2586 & 0.2903 & 0.3181 & 0.2929 & 0.2863 & 0.3232 & 0.3251 & 0.3258 & \textbf{0.3441} & $9.3e^{-10}$\\
&NDCG@40 & 0.0695 & 0.0833 & 0.0985 & 0.0994 & 0.1252 & 0.1390 & 0.1367 & 0.1532 & 0.1670 & 0.1555 & 0.1540 & 0.1745 & 0.1780 & 0.1751 & \textbf{0.1898} & $1.0e^{-9}$\\
\hline

\multirow{4}{*}{Yelp}
&Recall@20 & 0.0198 & 0.0304 & 0.0491 & 0.0510 & 0.0562 & 0.0584 & 0.0681 & 0.0742 & 0.0761 & 0.0700 & 0.0665 & 0.0806 & 0.0803 & 0.0789 & \textbf{0.0869} & $6.2e^{-7}$\\
&NDCG@20 & 0.0094 & 0.0143 & 0.0222 & 0.0245 & 0.0282 & 0.0280 & 0.0336 & 0.0365 & 0.0373 & 0.0347 & 0.0327 & 0.0402 & 0.0398 & 0.0391 & \textbf{0.0437} & $1.0e^{-6}$\\
\cline{2-18}
&Recall@40 & 0.0307 & 0.0487 & 0.0692 & 0.0743 & 0.0856 & 0.0891 & 0.1019 & 0.1151 & 0.1175 & 0.1072 & 0.1032 & 0.1230 & 0.1226 & 0.1210 & \textbf{0.1273} & $1.2e^{-4}$\\
&NDCG@40 & 0.0120 & 0.0187 & 0.0268 & 0.0315 & 0.0355 & 0.0360 & 0.0419 & 0.0466 & 0.0474 & 0.0437 & 0.0418 & 0.0505 & 0.0502 & 0.0492 & \textbf{0.0533} & $1.1e^{-5}$\\
\hline

\multirow{4}{*}{Amazon}
&Recall@20 & 0.0324 & 0.0367 & 0.0525 & 0.0486 & 0.0583 & 0.0837 & 0.0551 & 0.0772 & 0.0868 & 0.0617 & 0.0742 & 0.0955 & 0.0874 & 0.0885 & \textbf{0.1277} & $5.1e^{-13}$\\
&NDCG@20 & 0.0211 & 0.0234 & 0.0318 & 0.0317 & 0.0377 & 0.0579 & 0.0353 & 0.0501 & 0.0571 & 0.0372 & 0.0480 & 0.0623 & 0.5690 & 0.0578 & \textbf{0.0879} & $8.0e^{-13}$\\
\cline{2-18}
&Recall@40 & 0.0578 & 0.0600 & 0.0826 & 0.0773 & 0.0908 & 0.1196 & 0.0876 & 0.1175 & 0.1285 &0.0912 & 0.1123 & 0.1409 & 0.1312 & 0.1335 & \textbf{0.1782} & $7.3e^{-13}$\\
&NDCG@40 & 0.0293 & 0.0306 & 0.0415 & 0.0402 & 0.0478 & 0.0692 & 0.0454 & 0.0625 & 0.0697 & 0.0468 & 0.0598 & 0.0764 & 0.0704 & 0.0716 & \textbf{0.1048} & $1.5e^{-13}$\\
\hline
\end{tabular}
\vspace{-0.1in}
\label{tab:overall_performance}
\Description{A table presenting the evaluated performance of the proposed \model\ model and the baselines, in which \model\ significantly outperforms the baseline methods.}
\end{table*}

\vspace{-0.05in}
\subsubsection{\bf Baseline Methods}
For comprehensive performance comparision, our baseline set contains 14 recommendation methods.

\noindent\textbf{1) Conventional Collaborative Filtering Methods}
\begin{itemize}[leftmargin=*]
    \item \textbf{BiasMF}~\cite{koren2009matrix}: This is a widely-adopted matrix factorization baseline which projects users and items into latent vectors.
    \item \textbf{NCF}~\cite{he2017neural}: It is a representative neural CF method which replaces the inner-product with the non-linear feature projection.
\end{itemize}


\noindent\textbf{2) Autoencoder-based Recommender System}
\begin{itemize}[leftmargin=*]
    \item \textbf{AutoR}~\cite{sedhain2015autorec}: It employs the autoencoder to generate the embeddings of users and items with the reconstruction loss.
\end{itemize}


\noindent\textbf{3) GNN-based Collaborative Filtering}
\begin{itemize}[leftmargin=*]
\item \textbf{PinSage}~\cite{ying2018graph}: It leverages the graph convolution network to model the user-item interaction through graph structures.
\item \textbf{STGCN}~\cite{zhang2019star}: It design the graph autoencoding for high-order interaction learning with the reconstruction regularization.
\item \textbf{GCMC}~\cite{berg2017graph}: It is one of the pioneering CF models that enhances the matrix completion with GNN-based message passing.
\item \textbf{NGCF}~\cite{wang2019neural}: It is one of the state-of-the-art graph CF models that incorporates high-order connectivity for recommendation.
\item \textbf{GCCF}~\cite{chen2020revisiting} and \textbf{LightGCN}~\cite{he2020lightgcn}: To simplify the message passing scheme, recent studies propose to omit the non-linear feature transformation and activiation, which achieves better results.
\end{itemize}


\noindent\textbf{4) Disentangled Representation-enhanced GNN Model}
\begin{itemize}[leftmargin=*]
\item \textbf{DGCF}~\cite{wang2020disentangled}: This method disentangles latent factors behind user-item interactions under the graph neural network architecture, to capture the fine-grained user-item relationships.
\end{itemize}


\noindent\textbf{5) SOTA Self-Supervised Recommendation Methods}
\begin{itemize}[leftmargin=*]
    \item \textbf{SLRec}~\cite{yao2021self}: This self-supervised learning method conducts data augmentation with the learning of latent feature correlations.
    \item \textbf{NCL}~\cite{lin2022improving}: It enhances the graph contrastive learning with the augmented structural and semantic-relevant training pairs.
    \item \textbf{SGL}~\cite{wu2021self}: This method applies random structure augmentation such as node/edge drop and random walk, to generate contrastive views for graph augmentation in collaborative filtering.
    \item \textbf{HCCF}~\cite{xia2022hypergraph}: This method conducts local-global cross-view contrastive learning with hypergraph-based structure learning.
\end{itemize}



\vspace{-0.05in}
\subsubsection{\bf Hyperparameter Settings}
We implement \model\ with PyTorch. The Adam optimizer is used for parameter inference, with learning rate of $1e^{-3}$ and batch size of 4096. By default, the hidden dimensionality is set as 32, and the number of graph neural iterations is tuned from \{1,2,3\}. The graph self-attention employs 4 attention heads. We conduct subgraph sampling for every 10, 30, or 60 steps for further improving the model efficiency. In our learning to mask paradigm, the size of centric nodes is selected from $\{200, 400, ..., 1800, 2000\}$. The ratio of sampled nodes $\rho$ is set as 0.2. We select the number of graph hops for subgraph masking from \{1,2,3\} to reflect the high-order semantic relatedness. In the model training phase, $\lambda_1$ and $\lambda_2$ are respectively chosen from $\{1e^{-k}|-1\leq k\leq 4\}$ and $\{1e^{-k}|3\leq k\leq 8\}$ to control the regularization strength of weight-decay factor and the augmented self-supervised signals.


\vspace{-0.05in}
\subsection{Overall Performance Comparison (RQ1)}

\begin{figure}[t]
    \centering
    \subfigure[Gowalla-Recall]{
        \includegraphics[width=0.46\columnwidth]{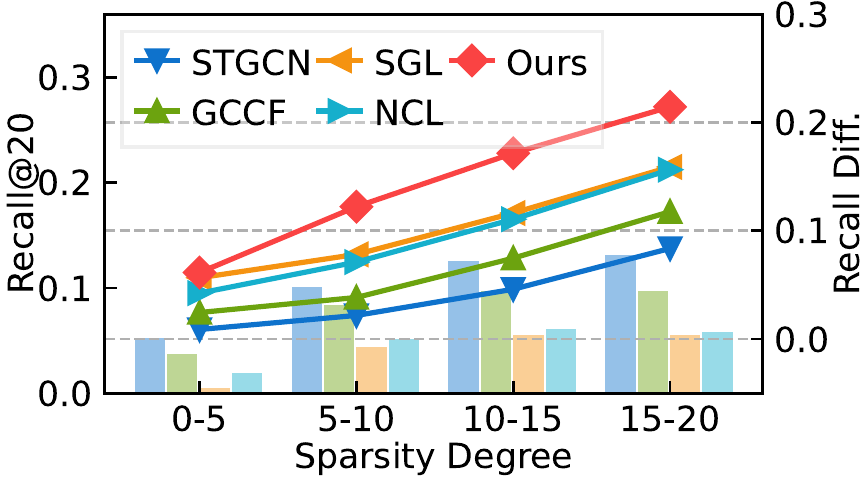}
    }
    \subfigure[Gowalla-NDCG]{
        \includegraphics[width=0.46\columnwidth]{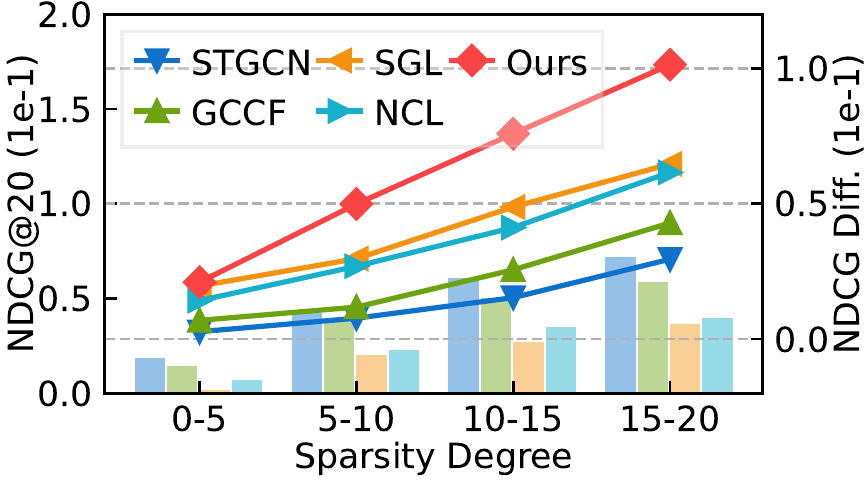}
    }
    \vspace{-0.1in}
    \caption{Performance against data sparsity. Left-side y-axis: performance curve of each method. Right-side y-axis: improvement ratio between \model\ and each baseline.}
    \label{fig:sparsity}
    \vspace{-0.1in}
    \Description{A line figure showing the performance of \model\ and baselines on sub-datasets with different sparsity degree, in which \model\ outperforms baselines in all sparsity degrees.}
\end{figure}

Table~\ref{tab:overall_performance} lists the overall performance of all compared methods on three datasets. From the results, we have the observations below:

\begin{itemize}[leftmargin=*]
\item \textbf{Obs.1}: \textbf{Superiority over SOTA SSL approaches}. \model\ consistently achieves the best performance compared with SOTA recommendation approaches by a large margin (measured by calculated $p$-values), including strong self-supervised approaches in all datasets. By incorporating the adaptive data augmentation into the graph SSL framework through our learning to mask paradigm, \model\ surpasses existing solutions with the automated self-supervised signal generation. The augmentation strategy in current SSL methods (\eg, SGL, HCCF) with self-discrimination over all nodes, which make them vulnerable to the perturbation, such as popularity bias and interaction noise. That is, with the careful design of our graph augmentation scheme adaptive for 
the infomax-based collaborative relation learning,
self-supervised signals with more beneficial instances will be automately identified for data augmentation. \\\vspace{-0.12in}


\item \textbf{Obs.2}: \textbf{Benefits with SSL augmentation}. Comparing various baselines, we notice that SSL methods (\ie, HCCF, SGL, NCL) outperforms\ pure GNN-based CF models (\eg, LightGCN, NGCF, GCCF) in most cases. This verifies the benefits of self-supervised data augmentation under the graph-based CF scenario. Such performance gap sheds light on the limitation of GNN-based CF model: i) easily aggregating noisy information for downstream recommendation task; ii) the over-fitting issue over sparse data, which weakens the model representation quality. \\\vspace{-0.12in}

\item \textbf{Obs.3}: \textbf{Performance against data sparsity}. We further investigate the recommendation performance of our \model\ and several representative baselines with respect to different data sparsity degrees. In particular, users are partitioned into four groups in terms of their interaction frequencies, \ie, $[0, 5), [5, 10), [10, 15), [15, 20)$. Bars in this figure indicate the performance improvement between our \model\ and each compared baseline corresponding to the ratio in the right side y-axis. The performance curves of all methods are shown in Figure~\ref{fig:sparsity} in terms of \emph{Recall@20} and \emph{NDCG@20}. The evaluation results demonstrate that our \model\ owns a consistent superiority over strong self-supervised learning baselines, which again confirms the effectiveness of our adaptive data augmentation in accurately learning user preference with small labeled interaction data. The data augmentation with stochastic operations, such as the random node/edge dropout in SGL and corrupt embedding with random noise perturbation in SLRec, may drop some important information for inactive users, making the data sparsity issue even worse. 
\end{itemize}

\begin{figure}[t]
    \vspace{-0.05in}
    \centering
    \subfigure[Yelp Data]{
        \includegraphics[width=0.22\columnwidth]{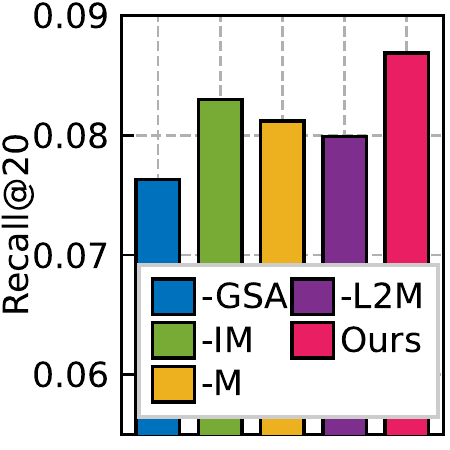}
        \includegraphics[width=0.22\columnwidth]{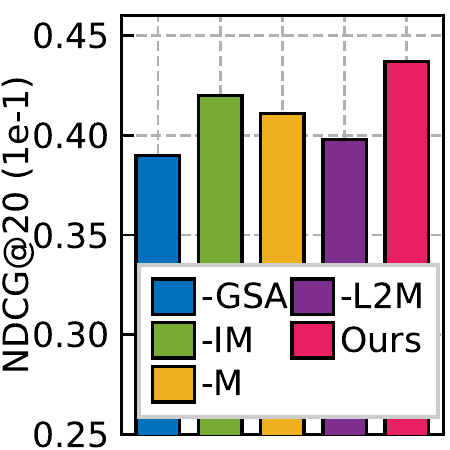}
    }
    \subfigure[Gowalla Data]{
        \includegraphics[width=0.22\columnwidth]{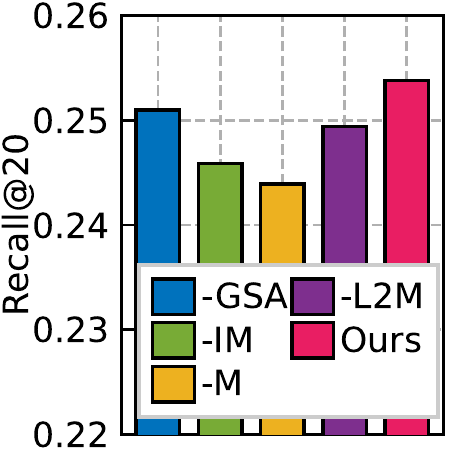}
        \includegraphics[width=0.22\columnwidth]{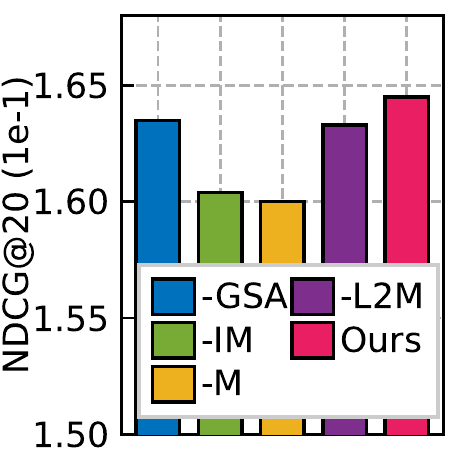}
    }
    \vspace{-0.1in}
    \caption{Module ablation study on Yelp and Gowalla.}
    \label{fig:ablation}
    \vspace{-0.1in}
    \Description{A bar figure showing the performance of four ablated models compared to the performance of \model. All ablated models perform worse in comparison.}
\end{figure}

\vspace{-0.1in}
\subsection{Module Ablation Study (RQ2)}
We conduct ablation study to investigate the individual contribution of different sub-modules of the proposed \model\ to the superior recommendation results. The results are reported in Figure~\ref{fig:ablation}.\\\vspace{-0.12in}



\noindent \textbf{Effect of graph self-attention}. We study the benefit of graph self-attention decoder for the reconstruction-based SSL task with the variant \textbf{-GSA}. In this variant, a symmetry encoder-decoder network is adopted to replace graph self-attention with the graph convolutional network. The results manifest that our \model\ has an obvious improvement over \textbf{-GSA}, which demontrates that aggregation with global self-attention over the augmented graph improves performance against the over-smoothing issue in GNNs.\\\vspace{-0.12in}

\noindent \textbf{Effect of reconstruction SSL signals}. We study the influence of our augmented generative self-supervised signals with the masked reconstruction objective. From the results of \model\ and variant \textbf{-M} (without the $\mathcal{L}_\text{InfoM}$ and $\mathcal{L}_\text{recon}$ regularization), the performance gain reveals that the design generative SSL task is beneficial for enhancing graph-based CF paradigm, by injecting auxiliary self-supervised signals to reflect true underlying patterns.\\\vspace{-0.12in}


\noindent \textbf{Effect of infomax-based semantic relatedness}.
The learnable mask generation function $\varphi(\mathcal{G}, \mathcal{V}; k)$ plays a critical role in the automated augmentation of our \model\ model by considering the infomax-based subgraph semantic relatedness $s_v$. Instead, the variant \textbf{-IM} removes the infomax-based loss $\mathcal{L}_\text{InfoM}$ in the mask generator to discard the guidance of local-global consistency. The performance degradation of \textbf{-IM} variant indicates that our \model\ can provide meaningful gradients to guide the model training in recovering importnat and noise-resistent collaborative relations. \\\vspace{-0.12in}


\noindent \textbf{Effect of learning to mask paradigm}. The adaptive centric node-based masking strategy ${\psi}'(v;k)$ is replaced with the randomly edge masking in the variant \textbf{-L2M}. While the SSL reconstruction task is still added for data augmentation, the non-adaptive augmention scheme may contain noisy information for suboptimal performance.

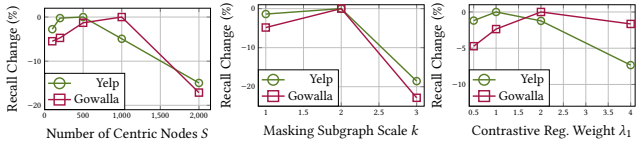
\begin{figure}[t]
    \centering
    \begin{adjustbox}{width=\columnwidth}
    \begin{tikzpicture}
\begin{axis}[
    width=3.2in,
    height=2.2in,
    xlabel={Number of Centric Nodes $S$},
    ylabel= {Recall Change (\%)},
    xmin=0,xmax=2150,
    ymin=-21,ymax=2,
    legend columns=1,
    legend cell align=right,
    grid=both,
    every axis plot/.append style={ultra thick},
    every tick label/.append style={scale=1.1},
    label style={scale=1.8},
    legend style={
        nodes={scale=1.5, transform shape},
        legend image post style={scale=1.5},
        },
    legend style={at={(0,0)},anchor=south west},
    every axis plot post/.append style={
        every mark/.append style={scale=2}
    }
]
\addplot[color={rgb:red,120;green,171;blue,54}, mark=o, mark options={solid}]
table[x=para, y=yelp_hr] {seedNum.txt};
\addplot[color={rgb:red,233;green,30;blue,99}, mark=square, mark options={solid}]
table[x=para, y=gowalla_hr] {seedNum.txt};
\legend{\large Yelp, \large Gowalla};
\end{axis}
\end{tikzpicture}

\begin{tikzpicture}
\begin{axis}[
    width=3.2in,
    height=2.2in,
    xlabel={Masking Subgraph Scale $k$},
    ylabel={Recall Change (\%)},
    xmin=0.9,xmax=3.1,
    ymin=-25,ymax=1,
    legend columns=1,
    legend cell align=right,
    grid=both,
    every axis plot/.append style={ultra thick},
    every tick label/.append style={scale=1.1},
    label style={scale=1.8},
    legend style={
        nodes={scale=1.5, transform shape},
        legend image post style={scale=1.5},
        },
    legend style={at={(0,0)},anchor=south west},
    every axis plot post/.append style={
        every mark/.append style={scale=2}
    }
]
\addplot[color={rgb:red,120;green,171;blue,54}, mark=o, mark options={solid}]
table[x=para, y=yelp_hr] {maskDepth.txt};
\addplot[color={rgb:red,233;green,30;blue,99}, mark=square, mark options={solid}]
table[x=para, y=gowalla_hr] {maskDepth.txt};
\legend{\large Yelp, \large Gowalla};
\end{axis}
\end{tikzpicture}

\begin{tikzpicture}
\begin{axis}[
    width=3.2in,
    height=2.2in,
    xlabel={Contrastive Reg. Weight $\lambda_1$},
    ylabel={Recall Change (\%)},
    xmin=0.4,xmax=4.1,
    ymin=-13,ymax=1,
    legend columns=1,
    legend cell align=right,
    grid=both,
    every axis plot/.append style={ultra thick},
    every tick label/.append style={scale=1.1},
    label style={scale=1.8},
    legend style={
        nodes={scale=1.5, transform shape},
        legend image post style={scale=1.5},
        },
    legend style={at={(0,0)},anchor=south west},
    every axis plot post/.append style={
        every mark/.append style={scale=2}
    }
]
\addplot[color={rgb:red,120;green,171;blue,54}, mark=o, mark options={solid}]
table[x=para, y=yelp_hr] {sslReg.txt};
\addplot[color={rgb:red,233;green,30;blue,99}, mark=square, mark options={solid}]
table[x=para, y=gowalla_hr] {sslReg.txt};
\legend{\large Yelp, \large Gowalla};
\end{axis}
\end{tikzpicture}
    \end{adjustbox}
    \vspace{-0.1in}
    \caption{Hyperparameter study for the proposed \model\ in terms of Recall@20 changes on Yelp and Gowalla datasets.}
    \vspace{-0.1in}
    \label{fig:hyperparam}
    \Description{A line figure showing the performance change with respect to the number of centric nodes, the masking subgraph scale, and the self-contrastive weight.}
\end{figure}

\vspace{-0.05in}
\subsection{Hyperparameter Investigation (RQ3)}
\label{sec:parameter_study}
This section investigates the effect of several important hyperparameters on the recommendation performance of \model. The evalution results in terms of \emph{Recall@20} and \emph{NDCG@20} are presented in Figure~\ref{fig:hyperparam}. Due to space limit, the y-axis represents the relative performance degradation ratio compared with the best accuracy. We summarize the results with the following observations:


\begin{itemize}[leftmargin=*]

    \item \textbf{Number of Centric Nodes} $S$: This hyperparameter determines the size of the identified centric nodes in our learning to mask paradigm. We vary parameter $S$ in the range of $200\leq S \leq 2000$. As shown in the results, we first observe the performance improvement trend by increasing the size of centric node set. This observation points out the positive effect of incorporating more self-supervised signals into our recommendation task. However, further increasing the number of centric nodes results in more masked interaction edges, which might lead to damaging to the graph structure and distort the user (item) representations. \\\vspace{-0.12in}

    
    \item \textbf{Masking Subgraph Scale} $k$: In our graph masked autoencoder, $k$ represents the maximum hops included in the masked subgraph. The larger $k$ value indicates higher-order node dependency modeling with respect to user (item) semantic relatedness. The best performance is achieved with 2 hops. By comparing with the adaptive masking within $1$-hop subgraph, the incorporation of $2$-order may bring benefits to the consideration of high-order semantic relatedness in our self-supervised augmentation. \\\vspace{-0.12in}
    
    
    \item \textbf{Contrastive Regularization Strength} $\lambda_1$: In our model training phase, $\lambda_1$ determines the strength of contrastive regularization to enhance the discrinimation ability of encoded representations. It can be seen that an approriate regularization weight improves the model representation ability against graph over-smoothing issue with better user preference discrinimation. 
    
    
\end{itemize}

\vspace{-0.05in}
\subsection{Model Efficiency Study (RQ4)}
\label{sec:model_efficiency}
\begin{figure}[t]
    \vspace{-0.1in}
    \centering
    \includegraphics[width=0.45\columnwidth]{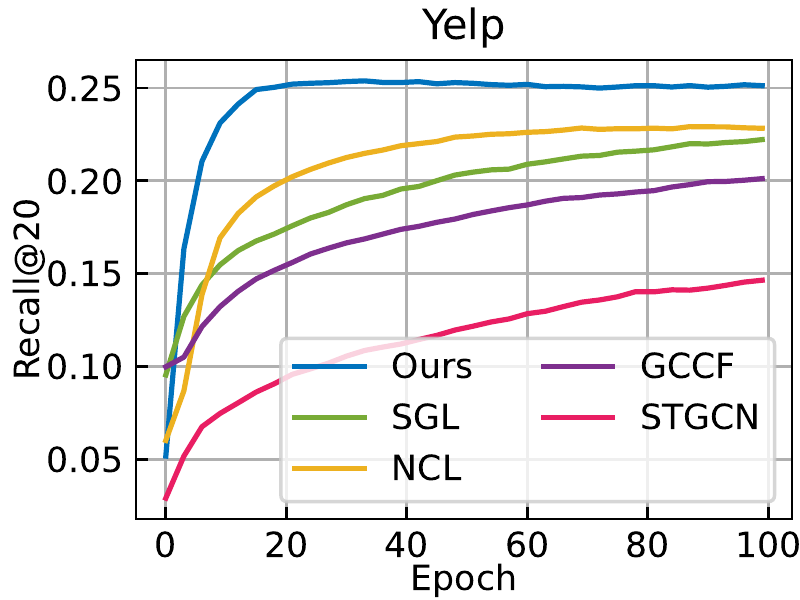}\quad
    \includegraphics[width=0.45\columnwidth]{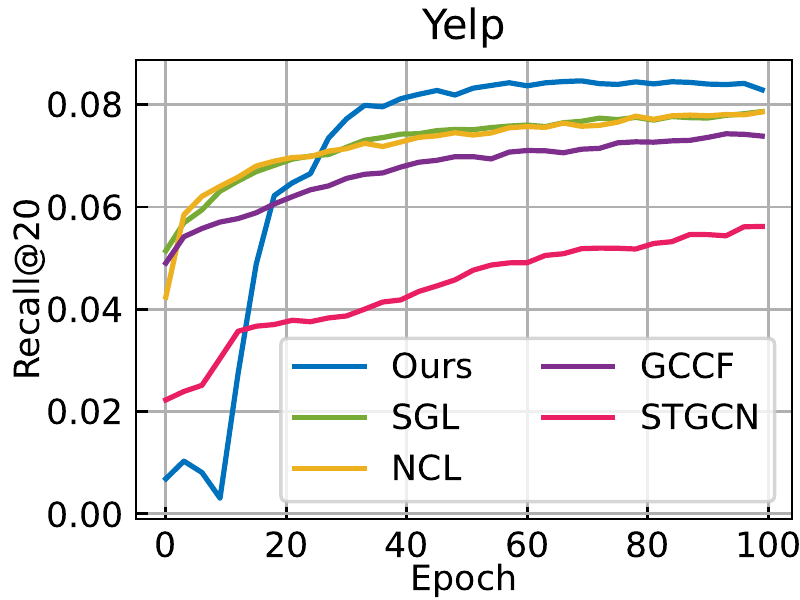}
    \vspace{-0.1in}
    \caption{Convergence analysis \wrt\ epochs while training.}
    \vspace{-0.05in}
    \label{fig:converge}
    \Description{A line figure showing the performance with respect to epochs for \model\ and some representative baselines. The figure shows that \model\ converges faster while training.}
\end{figure}

\begin{table}[]
    \centering
    \small
    \caption{Model efficiency study in terms of per-epoch training time (seconds) on Gowalla, Yelp, and Amazon datasets.}
    \label{tab:time}
    \vspace{-0.1in}
    \begin{tabular}{l|cccc|c}
        \hline
        Model & DGCF & NCL & HCCF & SGL & Ours\\
        \hline
        \hline
        Gowalla & 12.03s & 5.38s & 6.00s & 8.07s & 7.09s \\
        Yelp & 11.47s & 3.33s & 4.07s & 4.88s & 4.06s \\
        Amazon & 85.54s & 25.62s & 48.28s & 49.87s & 59.81s \\
        \hline
    \end{tabular}
    \vspace{-0.1in}
    \Description{A table showing the model efficiency in terms of per-epoch training time. \model\ is competitive in efficiency compared to the baselines.}
\end{table}

\subsubsection{\bf Model Convergence Analysis}
This section investigates the convergence of our \model\ and the results are depicted in Figure~\ref{fig:converge}. We observe that \model\ benefits from the incorporation of adaptive self-supervision signals for faster convergence to 20 and 40 training epochs on Gowalla and Yelp, respectively. While self-supervision is performed for data augmentation in state-of-the-art contrastive recommender systems (\ie, SGL, NCL), the weak robustness of those methods against noisy collaborative relations in graphs
, leads to their slower convergence compared with our \model\ model. This observation validates the training efficiency of \model, meanwhile maintaining superior performance. Such model advantage can be ascribed to that our offered SSL task with infomax-based semantic relatedness 
is helpful for guiding the optimization with better gradients.


\subsubsection{\bf Computational Cost Evaluation}
We further perform the model efficiency study and report the evaluation results in Table~\ref{tab:time}. While we enable the automated self-supervision signal distillation for data augmentation in recommender systems, we can still achieve competitive efficiency compared with state-of-the-art methods. Compared with our method, SGL performs the contrastive learning by constructing dual-view representation space for self-supervision, which requires more time for extra branch-based augmentation and graph-based embedding propagations.

We further evaluate the scalability of our method in handling large-scale datasets collected from an online retailer site. The evaluation results are presented in Supplementrary Section A.5.



\begin{figure}[t]
    \centering
    \includegraphics[width=0.94\columnwidth]{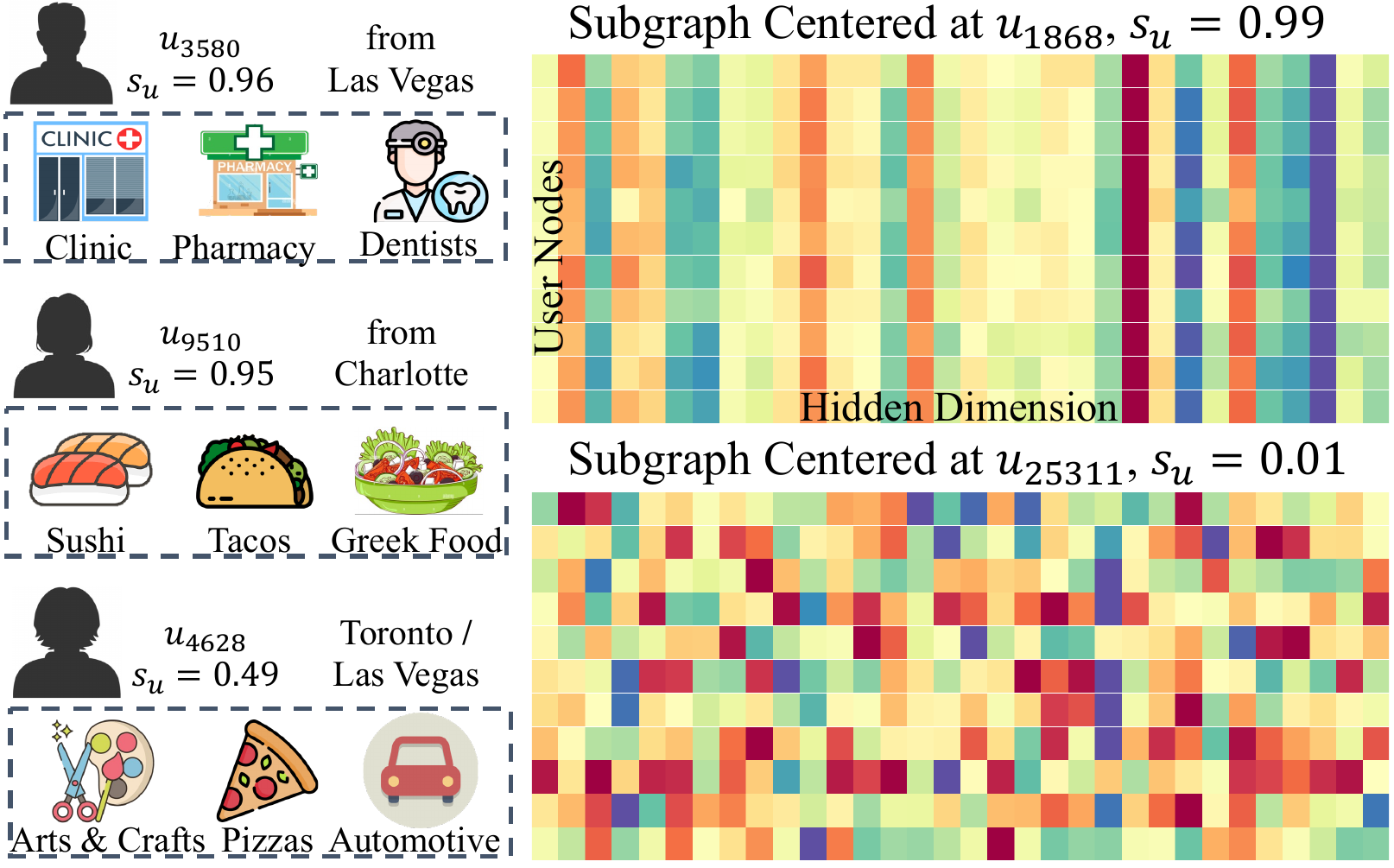}
    \vspace{-0.05in}
    \caption{Case study for i) model interpretation in learning semantic relatednes of user interactions; ii) heatmaps of encoded user embeddings from two user-centric subgraphs.}
    \label{fig:case_study}
    \vspace{-0.1in}
    \Description{A figure showing the learned local-global infomax scores and the related ground-truth information.}
\end{figure}

\subsection{Model Case Study (RQ5)}
\noindent{\bf Semantic Relatedness Interpretation}. In Figure~\ref{fig:case_study}, we sample three users ($u_{3580}$, $u_{9510}$, $u_{4628}$) with their learned subgraph semantic relatedness scores $s_v$. We see that $u_{4628}$ is observed to be more likely with interaction inconsistency compared with $u_{3580}$ and $u_{9510}$, because some of his/her interacted venues within a short time period are geographically located at distant cities (\eg, Las Vegas and Toronto). In contrast, $u_{3580}$'s interacted venues (\eg, Clinic, Dentists and Pharmacy) show high semantic relatedness in the same city. Additionally, we sample two subgraph from one identified centeric node ($u_{1868}$) and other node ($u_{25311}$). Each row represents the embedding of a user in the subgraph. By comparing the visualized node embeddings of these two subgraphs, we observe the strong homophily of the identified subgraph structure, which is consistent with high semantic relatedness for generative data augmentation. \\\vspace{-0.12in}

\begin{figure}[t]
    
    \centering
    \subfigure[HCCF]{
        \includegraphics[width=0.305\columnwidth]{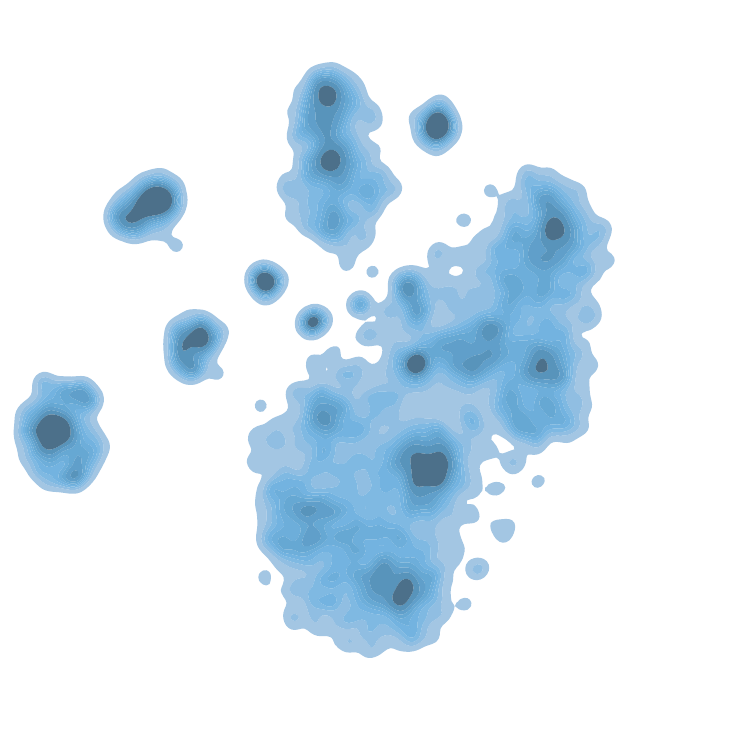}
    }
    \subfigure[NCL]{
        \includegraphics[width=0.305\columnwidth]{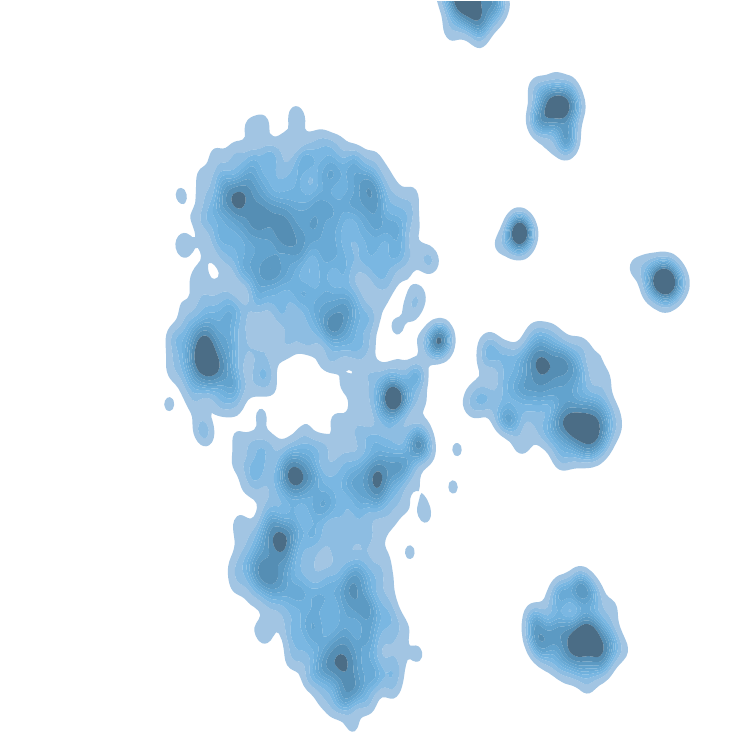}
    }
    \subfigure[Ours]{
        \includegraphics[width=0.305\columnwidth]{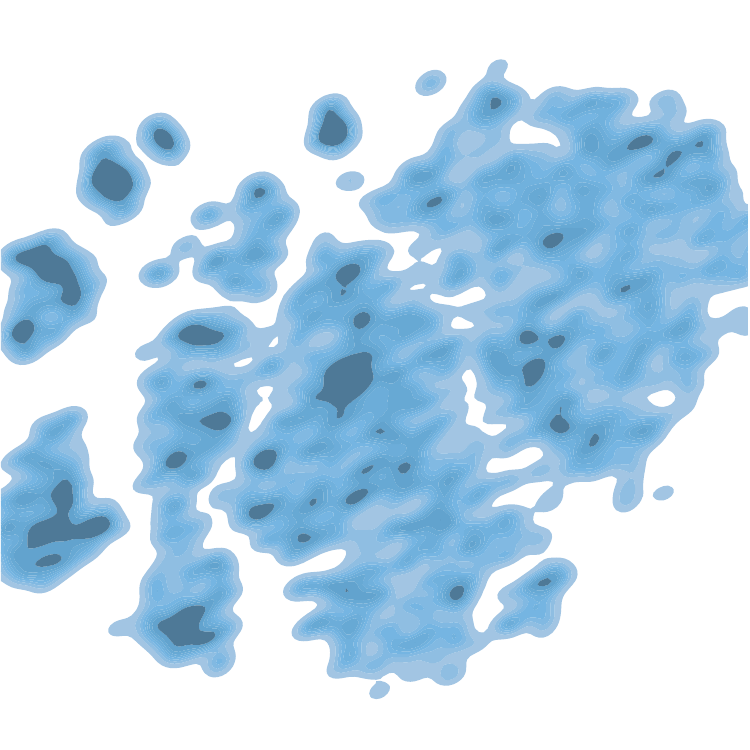}
    }
    \vspace{-0.1in}
    \caption{User embedding distribution of different methods with Gaussian kernel density estimation (KDE) on Yelp data.}
    \label{fig:my_label}
    \vspace{-0.1in}
    \Description{A figure showing the visualized embedding distribution learned by HCCF, NCL and the proposed \model, where the embedding given by \model\ spreads in a wider range.}
\end{figure}

\noindent{\bf Visualization of Representation Distributions}. We further show the distribution of user representations learned by different methods in 2-d space using tSNE and Gaussian kernel density estimation (KDE) in Figure~\ref{fig:my_label}. The embeddings encoded by \model\ exhibit better uniformity compared with SOTA contrastive approaches, which is indicative of better ability for preserving unique user preference.

\section{Related Work}
\label{sec:relate}

\noindent \textbf{GNN-based Recommender Systems}.
Due to the strength of GNNs, a line of research for recommendation focuses on enhancing the relational modeling with graph neural architectures in various recommendation scenarios~\cite{wu2020graph,gao2021graphrec}. For example, i) \emph{Social Recommendation}. GNN-based social recommender systems are developed to jointly perform message passing over the user-user and user-item connections, including GraphRec~\cite{fan2019graph}, KCGN~\cite{social2021knowledge} and ESRF~\cite{yu2020enhance}. ii) \emph{Sequential/Session-based Recommendation}. To improve the embedding quality, many recent attempts leverage GNNs to encode repreentations over the item sequences and capture the dynamic user preference, \eg, LESSR~\cite{chen2020handling}, GCE-GNN~\cite{wang2020global}, and SURGE~\cite{chang2021sequential}. iii) \emph{Knowledge graph-enhanced recommendation}. GNNs have also been adopted for improve the representation over the knowledge graphs to consider item semantic dependencies for recommendation, like KGAT~\cite{wang2019kgat}, KGNN-LS~\cite{wang2019knowledge} and KGCL~\cite{yang2022knowledge}. Towards this research line, our \model\ is built over GNN and advances the graph-based interaction modeling with automated data augmentation, which is less explored in existing solutions of recommender systems.\\\vspace{-0.12in}


\noindent \textbf{Self-Supervised Learning for Recommendation}. Recent studies bring the benefits of SSL into the recommender systems to address the data sparsity and noise challenges~\cite{wu2021self,wei2022contrastive,chen2022intent,xia2022self,xia2021self}. For example, contrastive SSL models have become the state-of-the-art recommendation paradigm by performing the data augmentation with their generated handcrafted or random contrastive views over graph structures, such as SGL~\cite{wu2021self}, HCCF~\cite{xia2022hypergraph} and NCL~\cite{lin2022improving}. In addition, some sequential models have also been improved with contrastive learning strategies to boost the recommendation performance, \eg, ICL~\cite{chen2022intent} and CL4SRec~\cite{xie2022contrastive}. However, the success of most contrastive learning-based recommender systems relies on manually designing effective contrastive views for reaching embedding agreement, which limits the model generalization ability. To fill this gap, this proposed \model\ method can offer great potential benefits for automated SSL signal distillation in recommendation. \\\vspace{-0.12in}

\noindent \textbf{Graph Autoencoders}. Autoencoder has become an effective neural network which is composed of the encoder to map input data into latent dimensions and the decoder to return the reconstruction of original input. It has inspired a wide range of graph learning applications, such as vertex feature reconstruction for node and graph classification~\cite{hou2022graphmae}, embedding reconstruction for graph-based recommendation~\cite{zhang2019star}, and structure reconstruction for molecular graph~\cite{ma2021gf}. Despite their effectiveness, most of existing graph autoencoders rely on the feature/structure reconstruction without masking or randomly masking, which can hardly be robust and adaptive to the target downstream graph learning tasks.


\section{Conclusion}
\label{sec:conclusoin}


The authors of this work have identified key limitations in existing self-supervised recommendation models, and propose a novel solution called the \model\ model. By incorporating self-supervised learning signals through infomax-based subgraph semantic relatedness, \model\ offers a graph augmentation scheme to distill informative self-supervision information in an automated manner. To evaluate the effectiveness of \model, the authors conducted extensive experiments on benchmark datasets and compared its performance to other types of recommender systems. The results demonstrate the effectiveness of our new approach compared with existing recommendation methods. This work highlights the potential benefits of automated self-supervised learning and provides an effective solution for enhancing recommender systems.




\section*{Acknowledgments}
This project is partially supported by Innovation and Technology Fund (Grant ITS/234/20) and 2022 Tencent Wechat Rhino-Bird Focused Research Program Research and Weixin Open Platform.

\clearpage
\bibliographystyle{ACM-Reference-Format}
\balance
\bibliography{sample-base}

\clearpage
\appendix \section{Appendix}
\balance
\label{sec:appendix}

In our provided appendix, we summarize the learning process of our \model\ recommender in Section A.1. Then, we present the implementation details of baseline methods in Section A.2. Additionally, we present the model robustness study against data noise and impact study of model depth in Section A.3 and Section A.4, respectively. Additionally, the model scalability investigation is presented in Section A.5. Finally, the derivation details of our \model\ model with therotical analysis are presented in Section A.6.

\subsection{Learning Algorithm Description}
We present the learning steps of our \model\ in Algorithm 1.
\begin{algorithm}[h]
	\caption{Learning Process of \model}
	\label{alg:learn_alg}
	\LinesNumbered
	\KwIn{User-item interaction graph $\mathcal{G}$, number of centric nodes in masking $S$, masking depth $k$, learning rate $\eta$, training epochs $E$, number of graph iterations $L$, regularization weights $\lambda_1, \lambda_2$.}
	\KwOut{Trained model parameters $\mathbf{\Theta}$.}
	Initialize model parameters $\mathbf{\Theta}$\\
	\For{$e=1$ to $E$}{
	    Initialize the maintained average difference of training loss by $\bar{\bigtriangledown} \mathcal{L}'_\text{rec}=0$\\
	    \For{mini-batch $\{(u,i)\}$ drawn from $\mathcal{E}$} {
	        Calculate the infomax-based semantic relatedness $s_v$ for each node $v$ (Eq~\ref{eq:infomax}) \\
	        Sample centric nodes $\mathcal{V}$ with Gumbel noise (Eq~\ref{eq:gumbel})\\
	        Conduct masking by $\mathcal{G}'=\varphi(\mathcal{G},\mathcal{V};k)$ (Eq~\ref{eq:masking})\\
	        Calculate the infomax-based loss $\mathcal{L}_\text{InfoM}$\\
	        \For{$l=1$ to $L$} {
	            Apply GCN encoding on $\mathcal{G}'$ (Eq~\ref{eq:gcn})\\
	        }
	        Sample edges $\tilde{\mathcal{G}}$ for graph self-attention (Eq~\ref{eq:gatSample})\\
	        Conduct graph self-attention decoding (Eq~\ref{eq:graphTrans})\\
	        Calculating the reconstruction loss $\mathcal{L}_\text{recon}$ (Eq~\ref{eq:pred})\\
	        Calculate the self-contrastive SSL loss $\mathcal{L}_\text{ssl}$ of the current batch (Eq~\ref{eq:ssl})\\
	        Calculate the recommendation loss and weight-decay regularization and combine them with the SSL loss to obtain $\mathcal{L}$ (Eq~\ref{eq:loss})\\
            \For{each parameter ${\theta}$ in $\mathbf{\Theta}$}{
                ${\theta} = {\theta} - \eta\cdot \frac{\partial \mathcal{L}}{\partial{\theta}}$;\\
            }
	    }
	}
    \Return all parameters $\mathbf{\Theta}$
    \Description{The algorithm for the learning process of \model.}
\end{algorithm}

\subsection{Baseline Implementation Details}
For fair performance comparison, we present the parameter setting details of some baseline methods. Specifically, the weight for weight-decay regularization is tuned from $\{1e^{-k}|3\leq k\leq 8\}$ for all baseline methods. The regularization strength of self-supervised learning task to supplement the main recommendation objective, is tuned from the suggested value range in compared SSL-based recommender systems. For example, the regularization weight for loss balance in NCL, HCCF, SGL, SLRec is tuned from $\{1e^{-k}|-1\leq k\leq 6\}$. The temperature parameter in the infoNCE loss for the contrastive learning-based baselines is tuned from $\{0.01, 0.03, 0.1, 0.3, 1, 3\}$. For baseline models that conduct edge or node dropout (\eg~LightGCN, SGL), the dropout rate is tuned from $\{0.1, 0.2, 0.3, 0.5, 0.8, 0.9\}$. For methods that were proposed for recommendation with explicit feedbacks (\eg~AutoRec, ST-GCN, GCMC), the models are trained using BPR ranking loss. For NCF, we evaluate the results given by the final NeuMF version. For ST-GCN, the reconstruction loss is tuned from $\{1e^{-k}|2\leq k\leq 6\}$. The number of intents in DGCF is tuned from $\{2, 4, 8\}$. For NCL, K-Means clustering is conducted in each $n$ epochs, $1\leq n\leq 5$.

\begin{figure}[h]
    \centering
    \vspace{-0.1in}
    \subfigure[Comparasion of different methods gainst noise data on Gowalla dataset]{
        \includegraphics[width=0.43\columnwidth]{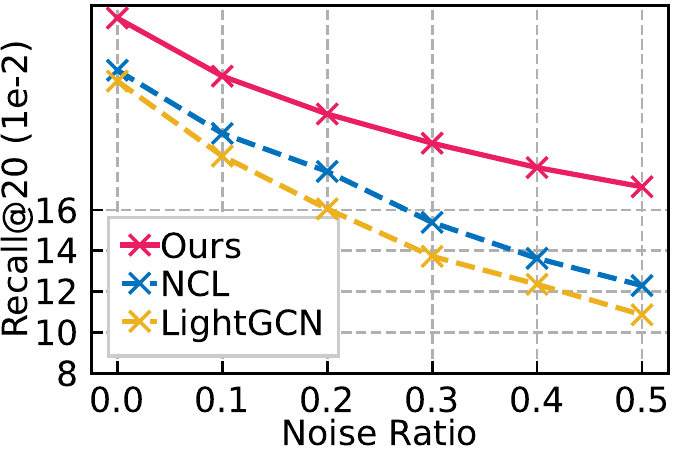}
        \quad
        \includegraphics[width=0.43\columnwidth]{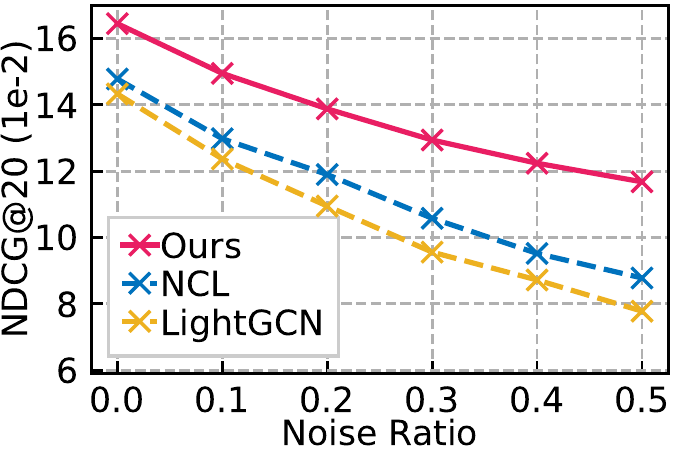}
    }
    \subfigure[Impact of data noise on Yelp dataset]{
        \includegraphics[width=0.43\columnwidth]{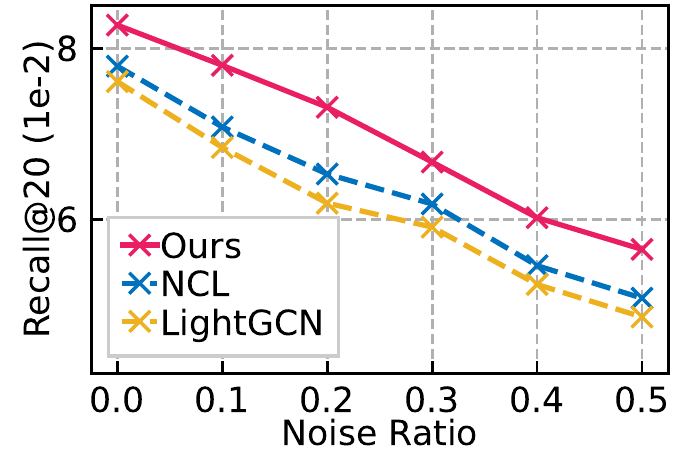}
        \quad
        \includegraphics[width=0.43\columnwidth]{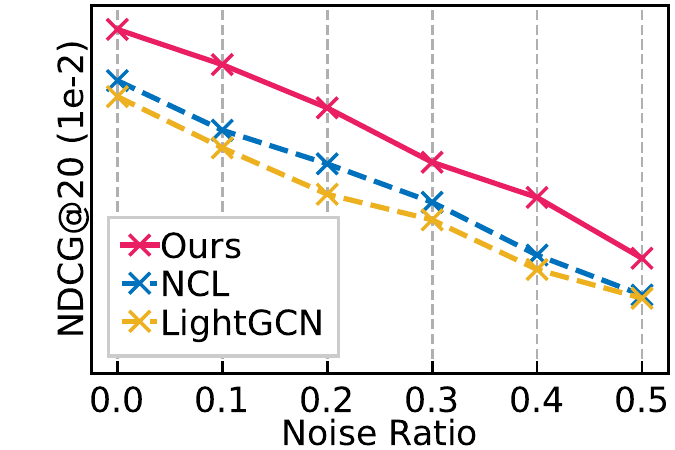}
    }
    \vspace{-0.2in}
    \caption{Full version of performance comparison in alleivating interaction noise in terms of Recall@20 and NDCG@20. x-axis represents the ratios of contaminated adversarial noisy user-item interactions in the training set. y-axis shows the performance of different methods against noise data.}
    \vspace{-0.15in}
    \label{fig:noise_origin}
    \Description{A figure showing the performance of \model\ and representative baselines with respect to noise ratio. \model\ consistently outperforms the baselines. Specially, \model\ is shown to be significantly more robust to high-ratio of noises.}
\end{figure}

\subsection{Model Robustness Study}
In addition to the presented results in Introduction section, we show the full version of performance comparison of different methods in Figure~\ref{fig:noise_origin} to study the impact of noisy data on performance. In particular, we corrupt the training set by randomly adding user-item interaction noisy examples with various percentages (\ie, from 10\% to 50\%). We keep the test set unchanged for all compared methods. The evaluation results show that our \model\ consistently outperforms state-of-the-art SSL recommender systems under different noise data ratios. It can be seen that our automated generative data augmentation is more beneficial for alleviating noise issue, which brings larger performance gain as the noise percentage increases. Even though current contrastive learning-based methods construct contrasting views for data augmentation, they may introduce noisy auxiliary self-supervision signals. Such noisy signals might mislead the data augmentation process. The superior performance of \model\ over compared methods, indicates the effectiveness of the designed learnable mask scheme to avoid involving noise in self-supervised augmentation, by considering the infomax-based semantic relatedness among collaborative relationships.

\begin{figure}[t]
    \centering
    \begin{adjustbox}{width=\columnwidth}
    \begin{tikzpicture}
\begin{axis}[
    width=3.2in,
    height=2.2in,
    xlabel={Number of Centric Nodes $S$},
    ylabel={NDCG Change (\%)},
    xmin=0,xmax=2150,
    ymin=-21,ymax=2,
    legend columns=1,
    legend cell align=right,
    grid=both,
    every axis plot/.append style={ultra thick},
    every tick label/.append style={scale=1.1},
    label style={scale=1.8},
    legend style={
        nodes={scale=1.5, transform shape},
        legend image post style={scale=1.5},
        },
    legend style={at={(0,0)},anchor=south west},
    every axis plot post/.append style={
        every mark/.append style={scale=2}
    }
]
\addplot[color={rgb:red,120;green,171;blue,54}, mark=o, mark options={solid}]
table[x=para, y=yelp_ndcg] {seedNum.txt};
\addplot[color={rgb:red,233;green,30;blue,99}, mark=square, mark options={solid}]
table[x=para, y=gowalla_ndcg] {seedNum.txt};
\legend{\large Yelp, \large Gowalla};
\end{axis}
\end{tikzpicture}

\begin{tikzpicture}
\begin{axis}[
    width=3.2in,
    height=2.2in,
    xlabel={Masking Subgraph Scale $k$},
    ylabel={NDCG Change (\%)},
    xmin=0.9,xmax=3.1,
    ymin=-25,ymax=1,
    legend columns=1,
    legend cell align=right,
    grid=both,
    every axis plot/.append style={ultra thick},
    every tick label/.append style={scale=1.1},
    label style={scale=1.8},
    legend style={
        nodes={scale=1.5, transform shape},
        legend image post style={scale=1.5},
        },
    legend style={at={(0,0)},anchor=south west},
    every axis plot post/.append style={
        every mark/.append style={scale=2}
    }
]
\addplot[color={rgb:red,120;green,171;blue,54}, mark=o, mark options={solid}]
table[x=para, y=yelp_ndcg] {maskDepth.txt};
\addplot[color={rgb:red,233;green,30;blue,99}, mark=square, mark options={solid}]
table[x=para, y=gowalla_ndcg] {maskDepth.txt};
\legend{\large Yelp, \large Gowalla};
\end{axis}
\end{tikzpicture}

\begin{tikzpicture}
\begin{axis}[
    width=3.2in,
    height=2.2in,
    xlabel={Contrastive Reg. Weight $\lambda_1$},
    ylabel={NDCG Change (\%)},
    xmin=0.4,xmax=4.1,
    ymin=-14,ymax=1,
    legend columns=1,
    legend cell align=right,
    grid=both,
    every axis plot/.append style={ultra thick},
    every tick label/.append style={scale=1.1},
    label style={scale=1.8},
    legend style={
        nodes={scale=1.5, transform shape},
        legend image post style={scale=1.5},
        },
    legend style={at={(0,0)},anchor=south west},
    every axis plot post/.append style={
        every mark/.append style={scale=2}
    }
]
\addplot[color={rgb:red,120;green,171;blue,54}, mark=o, mark options={solid}]
table[x=para, y=yelp_ndcg] {sslReg.txt};
\addplot[color={rgb:red,233;green,30;blue,99}, mark=square, mark options={solid}]
table[x=para, y=gowalla_ndcg] {sslReg.txt};
\legend{\large Yelp, \large Gowalla};
\end{axis}
\end{tikzpicture}
    \end{adjustbox}
    \vspace{-0.2in}
    \caption{Hyperparameter study for \model\ in terms of NDCG@20 changes on Yelp and Gowalla datasets.}
    \vspace{-0.1in}
    \label{fig:hyperparam_more}
    \Description{A line figure presenting the performance change of \model\ with respect to the number of centric nodes, the masking subgraph scale, and the self-contrastive weight.}
\end{figure}
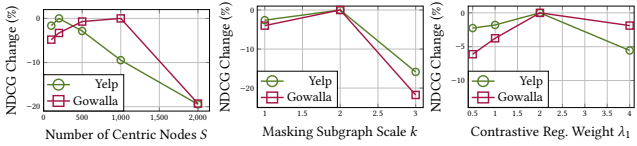

\begin{figure}[t]
    \centering
    \begin{adjustbox}{max width=1.0\linewidth}
    \begin{tikzpicture}
\begin{axis}[
    width=3.7in,
    height=2.2in,
    xlabel={\# Graph Neural Iterations $L$},
    ylabel={Recall Change (\%)},
    xmin=0.9,xmax=4.1,
    ymin=-50,ymax=3,
    legend columns=1,
    legend cell align=right,
    grid=both,
    every axis plot/.append style={ultra thick},
    every tick label/.append style={scale=1.1},
    label style={scale=1.8},
    legend style={
        nodes={scale=1.5, transform shape},
        legend image post style={scale=1.5},
        },
    legend style={at={(1,0)},anchor=south east},
    every axis plot post/.append style={
        every mark/.append style={scale=2}
    }
]
\addplot[color={rgb:red,120;green,171;blue,54}, mark=o, mark options={solid}]
table[x=para, y=yelp_hr] {gnnlayer.txt};
\addplot[color={rgb:red,233;green,30;blue,99}, mark=square, mark options={solid}]
table[x=para, y=gowalla_hr] {gnnlayer.txt};
\legend{\large Yelp, \large Gowalla};
\end{axis}
\end{tikzpicture}

\begin{tikzpicture}
\begin{axis}[
    width=3.7in,
    height=2.2in,
    xlabel={\# Graph Neural Iterations $L$},
    ylabel={NDCG Change (\%)},
    xmin=0.9,xmax=4.1,
    ymin=-50,ymax=3,
    legend columns=1,
    legend cell align=right,
    grid=both,
    every axis plot/.append style={ultra thick},
    every tick label/.append style={scale=1.1},
    label style={scale=1.8},
    legend style={
        nodes={scale=1.5, transform shape},
        legend image post style={scale=1.5},
        },
    legend style={at={(1,0)},anchor=south east},
    every axis plot post/.append style={
        every mark/.append style={scale=2}
    }
]
\addplot[color={rgb:red,120;green,171;blue,54}, mark=o, mark options={solid}]
table[x=para, y=yelp_ndcg] {gnnlayer.txt};
\addplot[color={rgb:red,233;green,30;blue,99}, mark=square, mark options={solid}]
table[x=para, y=gowalla_ndcg] {gnnlayer.txt};
\legend{\large Yelp, \large Gowalla};
\end{axis}
\end{tikzpicture}
    \end{adjustbox}
    \begin{adjustbox}{max width=1.0\linewidth}
    \begin{tikzpicture}
\begin{axis}[
    width=3.7in,
    height=2.2in,
    xlabel={Number of Latent Factors $d$},
    ylabel={Recall Change (\%)},
    xmin=6,xmax=66,
    ymin=-44,ymax=3,
    legend columns=1,
    legend cell align=right,
    grid=both,
    every axis plot/.append style={ultra thick},
    every tick label/.append style={scale=1.1},
    label style={scale=1.8},
    legend style={
        nodes={scale=1.5, transform shape},
        legend image post style={scale=1.5},
        },
    legend style={at={(1,0)},anchor=south east},
    every axis plot post/.append style={
        every mark/.append style={scale=2}
    }
]
\addplot[color={rgb:red,120;green,171;blue,54}, mark=o, mark options={solid}]
table[x=para, y=yelp_hr] {latFactor.txt};
\addplot[color={rgb:red,233;green,30;blue,99}, mark=square, mark options={solid}]
table[x=para, y=gowalla_hr] {latFactor.txt};
\legend{\large Yelp, \large Gowalla};
\end{axis}
\end{tikzpicture}

\begin{tikzpicture}
\begin{axis}[
    width=3.7in,
    height=2.2in,
    xlabel={Number of Latent Factors $d$},
    ylabel={NDCfG Change (\%)},
    xmin=6,xmax=66,
    ymin=-41,ymax=3,
    legend columns=1,
    legend cell align=right,
    grid=both,
    every axis plot/.append style={ultra thick},
    every tick label/.append style={scale=1.1},
    label style={scale=1.8},
    legend style={
        nodes={scale=1.5, transform shape},
        legend image post style={scale=1.5},
        },
    legend style={at={(1,0)},anchor=south east},
    every axis plot post/.append style={
        every mark/.append style={scale=2}
    }
]
\addplot[color={rgb:red,120;green,171;blue,54}, mark=o, mark options={solid}]
table[x=para, y=yelp_ndcg] {latFactor.txt};
\addplot[color={rgb:red,233;green,30;blue,99}, mark=square, mark options={solid}]
table[x=para, y=gowalla_ndcg] {latFactor.txt};
\legend{\large Yelp, \large Gowalla};
\end{axis}
\end{tikzpicture}
    \end{adjustbox}
    \vspace{-0.2in}
    \caption{Impact of model depth and dimensionality on the performance of \model, in terms of Recall@20 and NDCG@20 on both Yelp and Gowalla datasets.}
    \vspace{-0.1in}
    \label{fig:hyperparam_capacity}
    \Description{A line figure presenting the performance change of \model\ with respect to the number of graph neural iterations, and the number of latent factors.}
\end{figure}
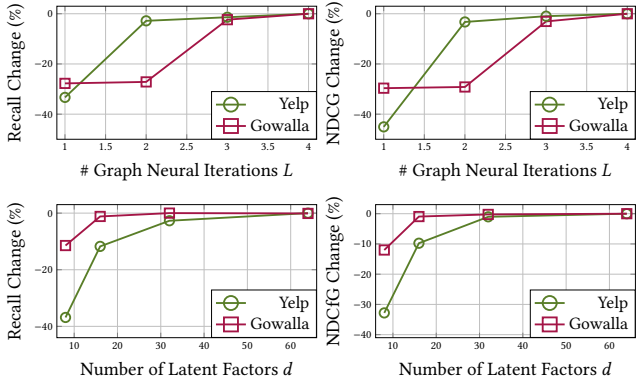

\subsection{Imapct of Depth and Dimensionality}
In this section, we present the results of impact study for model depth investigation and embedding dimensionality. From evaluation results in Figure~\ref{fig:hyperparam_capacity}, we have the following observations: 

\begin{itemize}[leftmargin=*]
    \item \textbf{Number of Graph Neural Iterations} $L$: This hyperparameter determines how many iterative graph neural network layers are applied in the masked autoencoder with graph self-attention. As shown in results, too few graph iterations result in insufficient graph connectivity modeling and damage the graph autoencoding learning. However, further increasing the graph iterations yields limited improvements due to oversmoothing.
    \item \textbf{Embedding Dimensionality} $d$: This is another general hyperparameter for model-based collaborative filtering. It controls the representation capacity of our \model. From the evaluation results, we observe typical underfitting-to-overfitting curves, where the performance increases dramatically with the increase of $d$ at the beginning (from $8$ to $16$ or $32$), and then the improvements become marginal when further increasing $d$ to $64$.
\end{itemize}

\noindent Figure~\ref{fig:hyperparam_more} presents the supplementary results for Section~\ref{sec:parameter_study}.

\subsection{Model Scalability Study}
\begin{figure}[t]
    \centering
    \includegraphics[width=0.45\columnwidth]{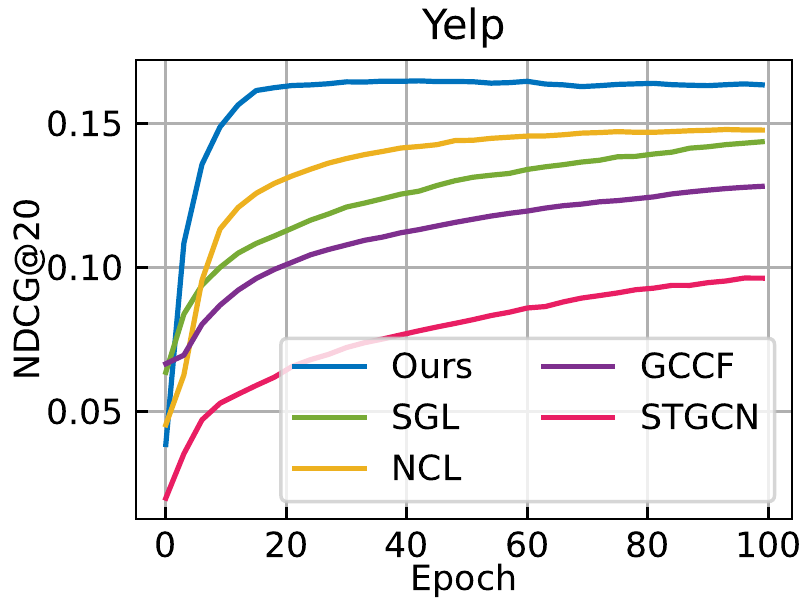}
    \quad
    \includegraphics[width=0.45\columnwidth]{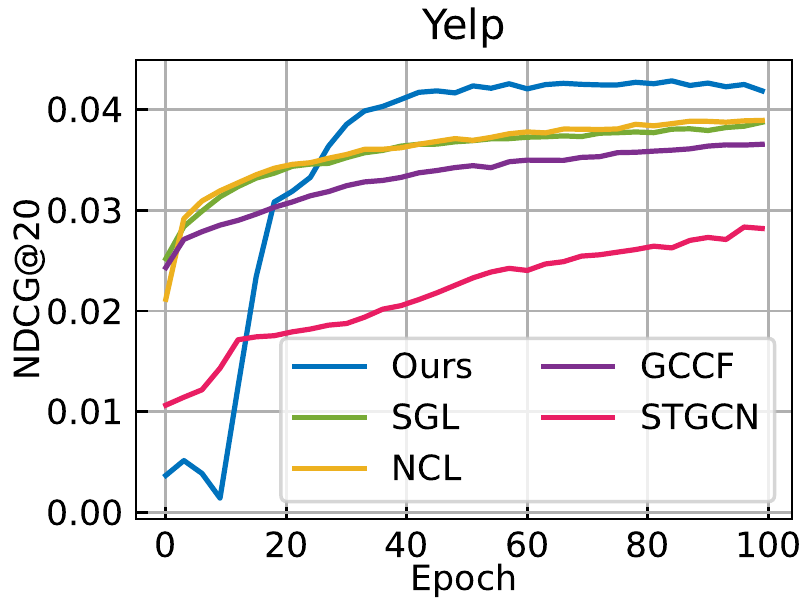}
    \vspace{-0.15in}
    \caption{Convergence analysis \wrt\ epochs while training on Gowalla and Yelp datasets, in terms of NDCG@20.}
    \label{fig:converge_more}
    \Description{A line figure showing the performance with respect to epochs for \model\ and some representative baselines. The figure shows that \model\ converges faster while training.}
\end{figure}

In this section, we report the running time of several methods in handling large-scale datasts (with millions of interactions) from an online retail platform. The system configuration for evaluation is Intel Xeon W-2133 CPU, NVIDIA TITAN RTX, with 64GB RAM. From results listed in Table~\ref{tab:scalability}, we can still observe that \model\ is able to achieve competitive efficiency compared with baselines, meanwhile maintaing much better recommendation performance. Convergence study in terms of NDCG@20 are presented in Figure~\ref{fig:converge_more} to supplement the results reported in Section~\ref{sec:model_efficiency}.

\begin{table}[t]
    \centering
    \footnotesize
    \setlength{\tabcolsep}{1.4mm}
    \caption{Model performance and per-epoch model training time of representative methods on large-scale Tmall dataset.}
    \label{tab:scalability}
    \vspace{-0.15in}
    \begin{tabular}{c|lccccc}
        \hline
        \# Edges & Metric & DGCF & SGL & HCCF & NCL & Ours\\
        \hline
        \hline
        \multirow{3}{*}{1.6M} & R@20 & 0.0221 & 0.0258 & 0.0272 & 0.0286 & \textbf{0.0299}\\
        & N@20 & 0.0258 & 0.0296 & 0.0309 & 0.0337 & \textbf{0.0354}\\
        \cline{2-7}
        & Time & 3569.5s & 1009.4s & 782.3s & 848.0s & 892.4s\\
        \hline
        \multirow{3}{*}{2.9M} & R@20 & 0.0253 & 0.0278 & 0.0283 & 0.0294 & \textbf{0.0306}\\
        & N@20 & 0.0279 & 0.0311 & 0.0319 & 0.0334 & \textbf{0.0358} \\
        \cline{2-7}
        & Time & 4828.6s & 1187.0s & 1040.5s & 1015.3s & 1158.8s\\
        \hline
    \end{tabular}
    \Description{A table showing the performance and the per-epoch training time of \model\ and baselines, on the large-scale Tmall data. \model\ is competitive in efficiency and outperforms the others.}
\end{table}

\subsection{Theoretical Analysis}
\label{sec:analysis}
We first give derivations on the gradients related to $\mathcal{L}_\text{recon}$. The reconstruction loss can be decomposed as follows:
\begin{align}
     -&\mathcal{L}_\text{recon}
    = \sum_{v_1,v_2} \hat{\textbf{h}}^\top_{v_1} \cdot \hat{\textbf{h}}_{v_2}
    = \sum_{v_1,v_2} (\textbf{h}_{v_1} + \tilde{\textbf{h}}_{v_1})^\top(\textbf{h}_{v_2} + \tilde{\textbf{h}}_{v_2})\nonumber\\
    &= \sum_{v_1, v_2} \textbf{h}_{v_1}^\top\textbf{h}_{v_2}+ \textbf{h}_{v_2}^\top \tilde{\textbf{h}}_{v_1}  + \textbf{h}_{v_1}^\top \tilde{\textbf{h}}_{v_2}  + \tilde{\textbf{h}}_{v_1}^{\top} \tilde{\textbf{h}}_{v_2}
\end{align}
For simplicity, here we only consider $v_1, v_2$ from the masked subgraph, which have no edges in the augmented graph and are thus not influenced by the graph convolutions. $\textbf{h}_{v_1}, \textbf{h}_{v_2}$ are randomly-initialized ego embeddings for node $v_1, v_2$, and $\tilde{\textbf{h}}_{v_1}, \tilde{\textbf{h}}_{v_2}$ are graph attention-based embeddings, which can be represented as:
\begin{align}
    \tilde{\textbf{h}}_{v} = \sum_{v'}\beta'_{v,v'} \textbf{W}'_\text{V} \textbf{h}_{v'}
\end{align}
where $\beta'_{v,v'}, \textbf{W}'_\text{V}$ denote the learned attention weights and the value transformation. Here we omit the multi-head setting for simplification. Then, the gradient of $\partial\mathcal{L}_\text{recon}/\partial\textbf{h}_{v_1}$ in Eq~\ref{eq:grad} is obtained.

Next, we show the derivations for the lower bound of the relatedness between a random pair of nodes $v_1$ and $v'$. Assuming the embedding vectors $\textbf{h}_{v}, \textbf{h}_{v_1}, \textbf{h}_{v'}$ are unit vectors, the pairwise distance can be calculated using a geometric trick, as follows:
\begin{align}
    d_{v,v_1} = 2\sin\frac{\theta_{v,v_1}}{2},~ d_{v,v'} = 2\sin\frac{\theta_{v,v'}}{2}, ~d_{v_1, v'} = 2\sin\frac{\theta_{v_1, v'}}{2}
\end{align}
where $\theta_{v, v_1},\theta_{v, v'}, \theta_{v_1, v'}\in[0, \pi]$ denote the angles between the corresponding vector pairs. As $v, v', v_1$ form a triangle (or in a line), the pairwise distances meet the following ristriction:
\begin{align}
    d_{v_1, v'} \leq d_{v,v_1} +d_{v,v'};~~~
    &\sin\frac{\theta_{v_1, v'}}{2} \leq \sin\frac{\theta_{v,v_1}}{2} + \sin\frac{\theta_{v,v'}}{2}\nonumber\\
    \sin^2\frac{\theta_{v_1, v'}}{2} \leq (\sin\frac{\theta_{v, v_1}}{2} + &\sin\frac{\theta_{v, v'}}{2})^2 < \sin^2\frac{\theta_{v, v_1}}{2} + \sin^2\frac{\theta_{v, v'}}{2}\nonumber\\
    1-\cos\theta_{v_1,v'} &<2 - \cos\theta_{v,v_1} - \cos\theta_{v,v'}\nonumber\\
    \cos\theta_{v_1,v'} &> \cos\theta_{v,v_1} + \cos\theta_{v,v'} - 1\nonumber\\
    \cos(v_1,v') &> \cos(v,v_1) + \cos(v,v') - 1
\end{align}

\end{document}